\documentclass[aps,prr,floatfix,preprintnumbers,altaffilletter,superscriptaddress,reprint,longbibliography,twocolumn]{revtex4-1}
\usepackage{graphicx,url,amssymb,amsmath,rotating,color,units,wasysym,epsfig,multirow,epstopdf}
\usepackage[colorlinks,urlcolor=blue,citecolor=blue,linkcolor=blue]{hyperref}
\usepackage{soul,xcolor}
\usepackage[printonlyused,nolist]{acronym}
\usepackage[ruled,vlined]{algorithm2e}
\usepackage{comment}

\newcommand{\new}{finite-duration}

\begin{document}

\title{Inference with finite time series: Observing the gravitational Universe through windows}

\author{Colm Talbot}
\email{colm.talbot@ligo.org}
\affiliation{LIGO Laboratory, California Institute of Technology, Pasadena, CA 91125, USA}
\affiliation{LIGO Laboratory, Massachusetts Institute of Technology, Cambridge, Massachusetts 02139, USA}
\affiliation{Kavli Institute for Astrophysics and Space Research, Massachusetts Institute of Technology, Cambridge, Massachusetts 02139, USA}

\author{Eric Thrane}
\affiliation{School of Physics and Astronomy, Monash University, VIC 3800, Australia}
\affiliation{OzGrav: The ARC Centre of Excellence for Gravitational-Wave Discovery, Clayton, VIC 3800, Australia}

\author{Sylvia Biscoveanu}
\affiliation{LIGO Laboratory, Massachusetts Institute of Technology, Cambridge, Massachusetts 02139, USA}

\author{Rory Smith}
\affiliation{School of Physics and Astronomy, Monash University, VIC 3800, Australia}
\affiliation{OzGrav: The ARC Centre of Excellence for Gravitational-Wave Discovery, Clayton, VIC 3800, Australia}

\begin{abstract}
Time series analysis is ubiquitous in many fields of science including gravitational-wave astronomy, where strain time series are analyzed to infer the nature of gravitational-wave sources, e.g., black holes and neutron stars.
It is common in gravitational-wave transient studies to apply a tapered window function to reduce the effects of spectral artifacts from the sharp edges of data segments.
We show that the conventional analysis of tapered data fails to take into account covariance between frequency bins, which arises for all finite time series---no matter the choice of window function.
We discuss the origin of this covariance and derive a framework that models the correlation induced by the window function.
We demonstrate this solution using both simulated Gaussian noise and real Advanced LIGO/Advanced Virgo data.
We show that the effect of these correlations is similar in scale to widely studied systematic errors, e.g., uncertainty in detector calibration and power spectral density estimation.
\end{abstract}

\maketitle

\section{Introduction}
Time-series analysis underpins recent advances in gravitational-wave astronomy.
The vast majority of gravitational-wave data analysis relies on windowing, a procedure that multiplies the time-domain data segment by a window function that tapers off at the beginning and end of the segment.
Analysts apply tapered windows to mitigate two effects: (1) spectral artifacts arising from the Fourier transform of the data segment edges (Gibbs phenomena) and (2) correlations between neighboring frequency bins.
While correlations between neighboring frequency bins can be reduced, they are never eliminated.

Choosing a suitable window requires balancing various considerations including the spectral leakage from instrumental lines and the low-frequency ``seismic wall,'' effectiveness mitigating the Gibbs phenomenon, and the loss of signal.
For a systematic study of the properties of different windows, we refer the reader to, e.g., ~\cite{Harris1978, Brillinger2001} for theoretical introductions, and~\cite{LIGODataAnalysis} for a specific discussion in the context of gravitational-wave data analysis.
Once the data are windowed, they are typically analyzed in the frequency domain where the noise is described by a \ac{PSD}, and it is assumed that each frequency bin is statistically independent.
However, this assumption is not true for a finite stretch of a longer noise process.

The assumptions underpinning the Whittle approximation have been thoroughly studied and many refinements have been proposed (e.g.,~\cite{Dahlhaus1988, Choudhuri2004, Rover2011, Sykulski2016, Kirch2017, Talbot2020, Rao2020}).
In this paper, we derive from first principles a formalism which accounts for the correlations between neighboring frequencies introduced by the window function applied to obtain finite time series.
We show how correlations between frequency bins arise from the fact that quasi-stationary Gaussian noise processes are fundamentally described in the frequency domain by continuous functions, which imply \textit{infinite-duration} time series.
We derive a simple expression for the ``\new'' covariance matrix, which encodes the correlations naturally present in all finite time series and identify our result as a specific basis for a \ac{KLT} (see, e.g.,~\cite{Maccone2009}).
We show that there are practical applications where the current conventional approach of windowing data incurs systematic errors, which though small, produce invalid inferences when data are combined in large sets or when we analyze gravitational-wave events with high \ac{SNR}.

The remainder of this paper is organized as follows.
In Section~\ref{sec:formalism}, we present the formalism underlying our framework.
We derive the \new{} covariance matrix for the analysis of finite time series. 
In Section~\ref{sec:demonstration}, we perform a demonstration, applying our method to the binary black hole merger events, GW150914~\cite{GW150914}, GW170814~\cite{GW170814}, and GW190521~\cite{GW190521} and contrast with results neglecting covariances.
We show how the current conventional windowing procedure can lead to faulty inferences when many gravitational-wave measurements are combined.
While our demonstration uses data from gravitational-wave astronomy, the framework we put forward is broadly applicable to all time-domain analysis.
We show that the problem is fixed by using the \new{} covariance matrix.
We provide closing thoughts in Section~\ref{sec:conclusions}.

\section{Formalism}\label{sec:formalism}
In this section, we derive a likelihood that enables us to analyze time-series data characterized by stationary, Gaussian noise in a way that correctly takes into account covariance between neighboring frequency bins that arises for all finite time series.

\subsection{Basic notation}
We consider time-series data $d(t)$ consisting of signal $s(t)$ and noise $n(t)$
\begin{align}
    d(t) = s(t) + n(t) .
\end{align}
In gravitational-wave observatories like LIGO~\cite{aLIGO} and Virgo~\cite{aVirgo}, $n(t)$ is a time series of dimensionless strain (change in length per unit length).
Transient gravitational-wave signals from merging binaries are characterized by comparing the data to gravitational waveform templates $h(t)$.

\subsection{Continuous, infinite-duration noise}
To start, we focus on noise in the absence of signals.
The noise can be expressed in the frequency domain as
\begin{align}
    \tilde{n}(f) = \int^{\infty}_{-\infty} dt \, e^{-2\pi i ft} n(t) .
\end{align}
The noise is best described as continuous because it is defined for an arbitrary choice of frequency: with a sufficiently long measurement, it is possible \textit{in principle} to achieve sufficient resolution to measure $\tilde{n}(f)$ for any value of $f$.

If we assume the noise is Gaussian, the likelihood of observing a specific noise realization is characterized by a covariance matrix
\begin{align}\label{eq:Cab}
    \mathcal{C}_{\mu\nu} &= \frac{1}{2} \left< \tilde{n}(f_\mu) \tilde{n}^{*}(f_\nu) \right>
\end{align}
the diagonal of which is equal to the noise \ac{PSD} $\mathcal{S}$ 
\begin{equation}
    \mathcal{S}_{\mu} = {\rm diag}\left( \mathcal{C}_{\mu\nu} \right) .
\label{eq:PSD}
\end{equation}

We refer to $\mathcal{C}_{\mu\nu}$ as the ``infinite-duration'' covariance matrix.
It is defined \textit{continuously} for arbitrary values of $f_\mu$ and $f_\nu$ and, in the time domain, it is defined for all times: $(-\infty, +\infty)$.
Throughout, repeated indices are summed over unless otherwise specified.
In the next subsection, we contrast $\mathcal{C}_{\mu\nu}$ (calligraphic script and greek indices) with the \new{} covariance matrix, denoted $C_{ij}$ (no calligraphic script and roman indices), which is defined only for discrete frequency bins $f_i$ and $f_j$ (or equivalently, for a finite duration).
If we further assume that the noise is stationary (the \ac{PSD} does not vary in time), $\mathcal{C_{\mu\nu}}$ is diagonal.

We note that, even if the true covariance matrix is diagonal, a naive empirical estimate of this quantity necessarily has some statistical uncertainty and will not generically be diagonal, even for a stationary Gaussian process.
We emphasize that in this section we seek to derive expressions for the true noise covariance matrix assuming $\mathcal{C}_{\mu\nu}$ is known.
and neglect the impact of empirical estimates.
In Appendix~\ref{app:infinite-duration}, we describe how we estimate $\mathcal{C}_{\mu\nu}$ in practice.

\subsection{Non-continuous, finite-duration noise}

In practice, we only consider finite stretches of data.
In this subsection, we derive the properties of finite stretches of continuous noise.
Let us consider data measured with sampling rate $f_s$ over data segment duration $T$.
There are 
\begin{align}
    N = f_s T
\end{align}
independent measurements.
We assume that the noise has no content above half the sampling rate and so we can probe every frequency without aliasing.
In practice, applying an aggressive low-pass filter removes this high-frequency content.

These data can be represented either in the time-domain as a real $N$-component time series with spacing $1/f_s$ or in the frequency domain as a complex frequency series $\tilde{d}_i$ 
with $-f_s / 2 \leq f \leq f_s / 2$ with spacing $1/T$ where the endpoints and zero-frequency component are required to be real.
These two domains are related via the discrete Fourier transform~\footnote{We note that here we use two-sided discrete Fourier transforms rather than the one-sided version widely used in gravitational-wave data analysis.}
\begin{equation}\label{eq:fft}
    \tilde{d}_k' = \frac{1}{f_s} \sum_{j=0}^{N - 1} d_j' e^{-2\pi i jk/N} .
\end{equation}
The frequency-domain covariance matrix for finite-duration, non-continuous noise is:
\begin{equation}\label{eq:average-covariance}
    C_{ij} = \frac{1}{2} \left< \tilde{d}'_i \tilde{d}^{*\prime}_j \right>.
\end{equation}
Here, the angled brackets denote ensemble averages over noise realizations.
The widely-used Whittle approximation assumes that data at each of the analyzed frequencies are independent, i.e., $C_{ij}$ is a diagonal matrix.
This is generally a good approximation.
However, as we show in this paper, the assumption of independence is not strictly valid when analyzing a finite stretch of data, especially when using a tapered window.

We begin by defining our window function $w$, which describes how we measure some segment of noise from what is, in theory, an infinite-duration noise process:
\begin{align}
    \tilde{d}_{k}'
    &= \frac{1}{f_s} \sum_{\psi=-\infty}^{\infty} d_\psi w_\psi e^{-2\pi i \psi k / N} \\
    &= \frac{1}{f_s} \sum_{j=0}^{N} d_j w_j e^{-2\pi i j k / N} \\
    &= (\tilde{d} * \tilde{w})_k .
\end{align}
Here, $w_j$ is a time-domain window function and the frequency-domain noise is now the convolution of the original frequency-domain noise with the Fourier transformed window function.
The prime denotes quantities associated with the windowed data.

We stress that this window function is always present in gravitational-wave data analysis problems and is defined for all times, not just the analysis segment.
It is often ignored when it is a top hat function, i.e.,
\begin{equation}
    w_{\psi} = \begin{cases}
    1 &\quad 0 \leq \psi < N \\
    0 &\quad \text{else}
    \end{cases}.
\end{equation}
The most commonly used window function in parameter estimation for compact binary coalescences is the Tukey window
\begin{equation}
    w_{\psi}(\alpha) = \begin{cases}
    \frac{1}{2}\left[1 - \cos\left( \frac{2\pi \psi}{\alpha N} \right) \right] &\quad 0 < \psi < \frac{\alpha N}{2} \\
    1 &\quad \frac{\alpha N}{2} \leq \psi \leq N - \frac{\alpha N}{2} \\
    \frac{1}{2}\left[1 - \cos\left( \frac{2\pi (N - \psi)}{\alpha N} \right) \right] &\quad N - \frac{\alpha N}{2} < \psi < N \\
    0 &\quad \text{else}
    \end{cases}.
\label{eq:inifite-tukey}
\end{equation}
Common limiting cases of the Tukey window are the rectangular window ($\alpha=0$) and the Hann window ($\alpha=1$).
Throughout this paper, we use Tukey windows with $\alpha=0.1$ unless otherwise specified, although the formalism described here holds for arbitrary window functions.

Since convolution is a linear operation, we can express the windowed frequency-domain data using standard linear algebra notation
\begin{equation}\label{eq:data-project}
    \tilde{d}'_{k} = \tilde{W}_{k\mu} \tilde{d}_\mu ,
\end{equation}
where repeated indices denote summation.
Here, $\tilde{W}_{k\mu}$ is a non-square subset of the circulant matrix
\begin{equation}
    \tilde{W}_{\mu\nu} = \tilde{w}_{\mu - \nu}
\end{equation}
that projects infinite-duration noise with frequency resolution $\delta f \rightarrow 0$ to finite duration data with frequency resolution $1 / T$.
Here, $\tilde{w}$ is the discrete Fourier transform of the time-domain window function.

\begin{figure}
    \centering
    \includegraphics[width=\linewidth]{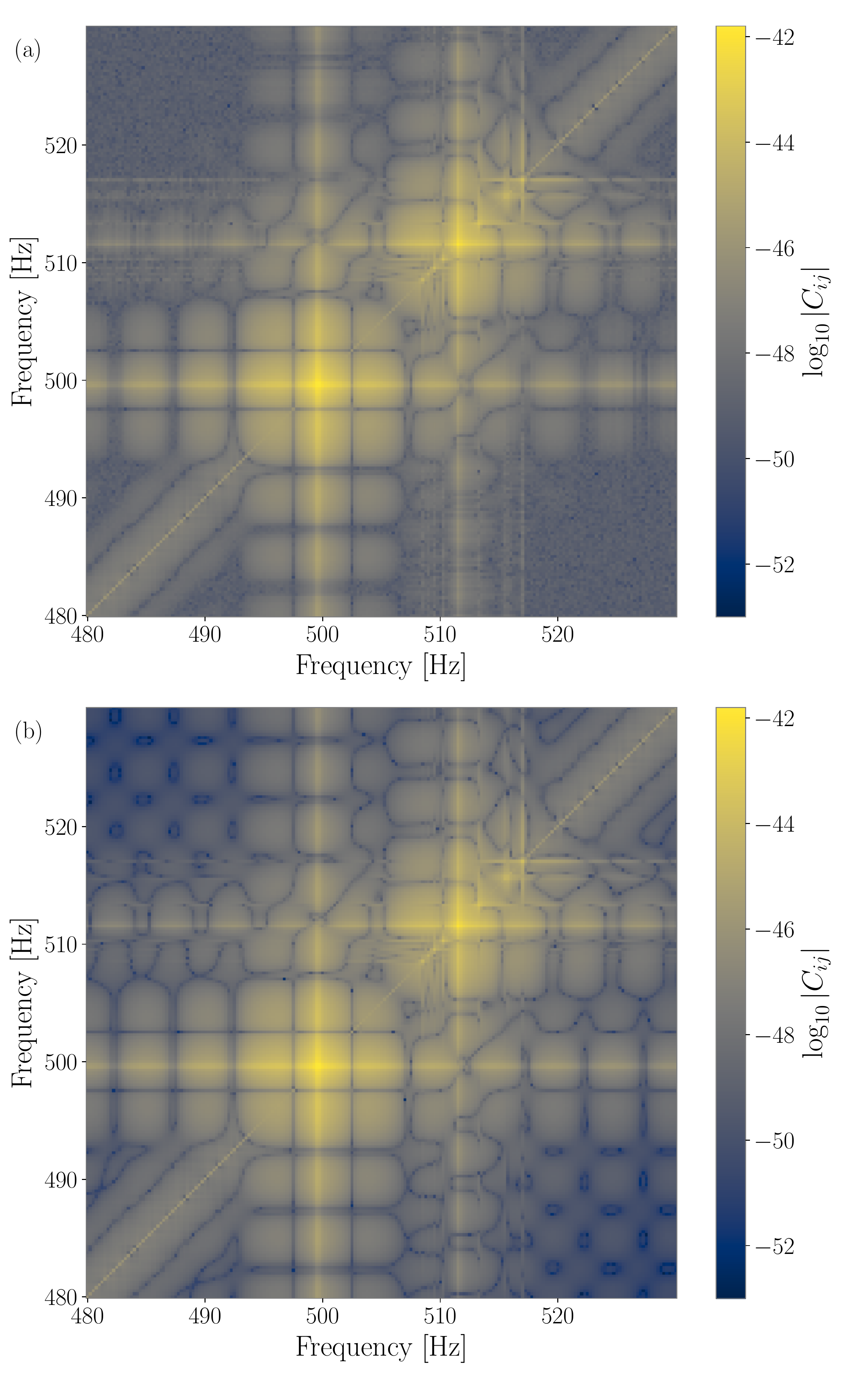}
    \caption{
    The estimated (top) and analytic (bottom) covariance matrix for simulated noise using a noise power spectral density estimated from LIGO Livingston data around the time of the binary black hole merger GW170814.
    The color bar indicates $\log_{10}$ power spectral density in units of $\unit[]{Hz^{-1}}$.
    The estimate is obtained using approximately three months of simulated data.
    The off-diagonal behavior agrees well with the analytic expression.
    The window used is a Tukey window with $\alpha=0.1$.
    }
    \label{fig:matrices}
\end{figure}

We can now write the covariance matrix for a finite-duration data stream with an arbitrary window function in terms of the frequency-domain covariance matrix of the infinite-duration process and the window function using Eq.~\ref{eq:average-covariance} and Eq.~\ref{eq:data-project}
\begin{align}
    C_{ij} &= \frac{1}{2} \left< \tilde{W}_{i\mu} \tilde{W}^{*}_{j\nu} \tilde{d}_\mu \tilde{d}^{*}_\nu \right> \\
    &= \frac{1}{2} \tilde{W}_{i\mu} \tilde{W}^{*}_{j\nu} \left< \tilde{d}_\mu \tilde{d}^{*}_\nu \right> \nonumber \\
    &= \tilde{W}_{i\mu} \tilde{W}^{*}_{j\nu} \mathcal{C}_{\mu\nu} .
\label{eq:exact-covariance}
\end{align}
If the underlying data are Gaussian and stationary, $\mathcal{C}_{\mu\nu}$ is diagonal and the finite-duration covariance matrix only depends on the window function and the infinite-duration \ac{PSD}.
While we initially defined the roman indices as covering frequencies from $[- f_{s} / 2, f_{s} / 2]$, in practice we analyze a narrower (positive) frequency range from $[f_{\min}, f_{\max}]$.
In this paper we will set $f_{\min}=20$ Hz, $f_{\max}=800$ Hz; this omits data that are affected by the bandpass filter we apply.
From here, roman indices will refer to this frequency range only.

Formally, one must carry out matrix products over the infinite axes denoted by greek indices to obtain the finite-duration covariance matrix in Eq.~\ref{eq:exact-covariance}.
In practice, however, via numerical experiment (see Appendix~\ref{app:infinite-duration}) we find that infinite-duration matrices can be adequately approximated using a frequency resolution 16 times that of the analysis segment.
In other words, when analyzing a $\unit[4]{s}$ data segment (frequency resolution = $\unit[0.25]{Hz}$), we may model infinite-duration noise with a $\unit[1 / 64]{Hz}$ (or higher-resolution) noise model.
In practice, we use a $\unit[1 / 128]{Hz}$ resolution.
The resolution of the noise model can be tuned to achieve the necessary precision for a given problem.

In Fig.~\ref{fig:matrices} we show the empirical finite-duration covariance matrix in the neighborhood of the diagonal---obtained by averaging approximately three months of simulated Gaussian noise broken into $10^{6}$ $\unit[4]{s}$ segments and using a Tukey window with $\alpha=0.1$ (Eq.~\ref{eq:average-covariance}) (top panel)---compared with the exact finite-duration covariance matrix obtained with our analytic expressions (Eq.~\ref{eq:exact-covariance}) (bottom panel).
The underlying \ac{PSD} ($\mathcal{S}_{\mu}$) is estimated using data from the LIGO Livingston interferometer at the time of the binary black hole merger GW170814 with a resolution of $\unit[1 / 128]{Hz}$ using the method described in Appendix~\ref{app:infinite-duration}.
The two panels agree well, demonstrating the correctness of this formalism.
However, the averaging estimate converges unacceptably slowly for practical use, and so our analytic expression is essential for practical applications.

We are interested in the inverse of the covariance matrix $C^{-1}_{ij}$.
We discuss the issue of inverting this matrix in Sec.~\ref{sec:inversion}.
We also emphasize that the finite-duration PSD (the leading diagonal of $C_{ij}$) $S_i \neq \mathcal{S}_{i}$, i.e., the finite-duration PSD is not the infinite-duration evaluated at the desired frequencies.
Additional technical details about this formalism are provided in the Appendix.
In Appendix~\ref{appendix:coarse-grain}, we discuss how our formalism is related to ``coarse-graining'' procedures used for \ac{PSD} estimation (e.g.,~\cite{S6Isotropic}).
In Appendix~\ref{computing} and Algorithm~\ref{algo:psd-calculation}, we describe in detail how we approximate $\mathcal{C}_{\mu\nu}$ and $C_{ij}$ from real data.

\subsection{Quantifying spectral leakage}

\begin{figure*}
    \centering
    \includegraphics[width=\linewidth]{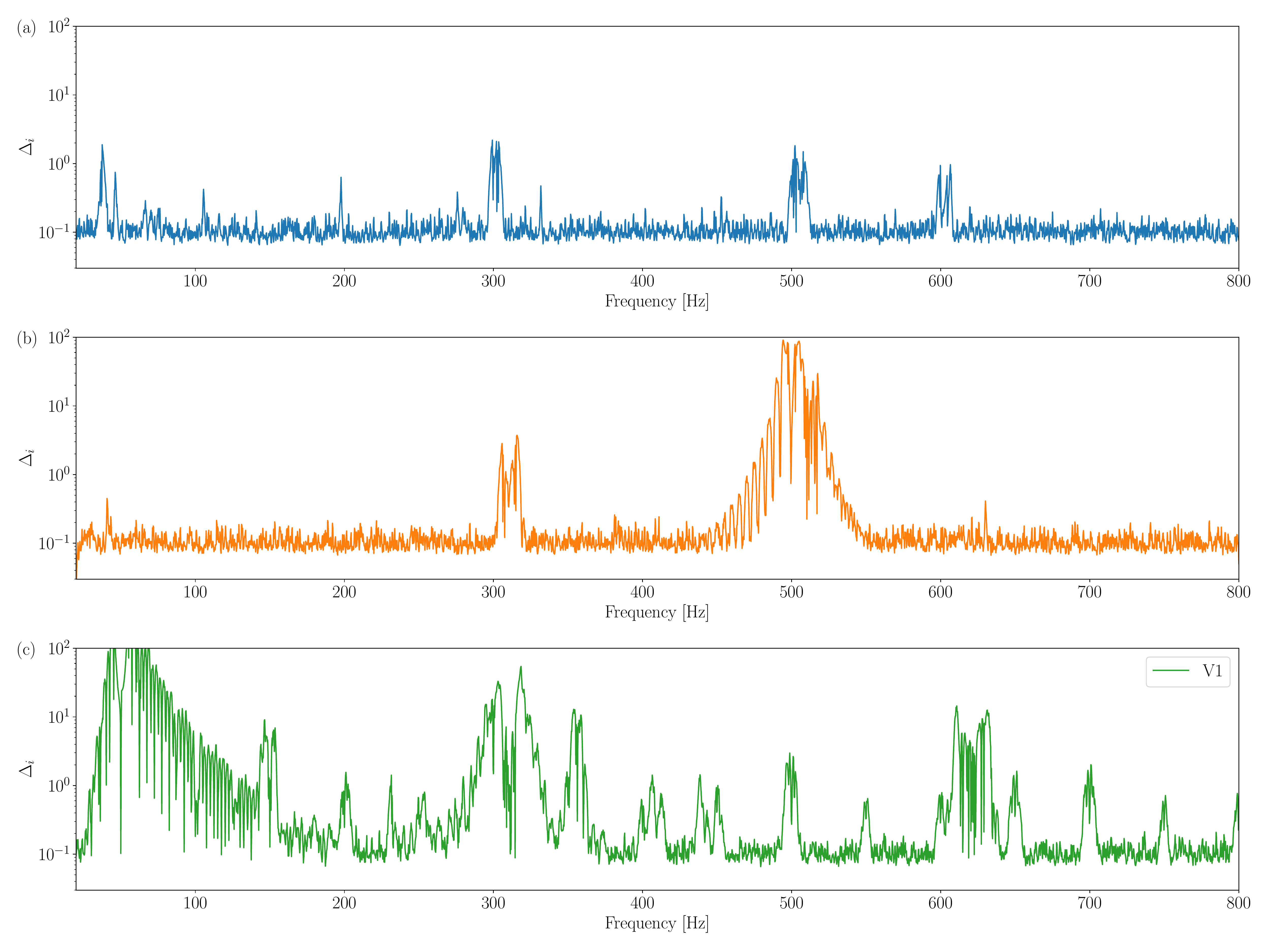}
    \caption{Maximum contamination per frequency bin (Eq.~\ref{eq:contamination}) for \ac{PSD}s estimated near the time of binary black hole merger GW170814~\cite{GW170814}.
    We see two competing effects.
    First, there is broadband contamination at $\sim 10\%$ when the \ac{PSD} is slowly varying.
    The magnitude of this contamination rises with increasing Tukey $\alpha$ (see Appendix~\ref{app:window-overlaps}).
    There is also large contamination near instrumental lines due to spectral leakage that is suppressed with increasing Tukey $\alpha$.
    }
    \label{fig:max-off-diagonal}
\end{figure*}

From Fig.~\ref{fig:matrices}, we see that windowing produces off-diagonal elements in the finite-duration covariance matrix in general.
In this subsection, we illustrate how this covariance is compounded by sharp spectral features.
The relationship between the choice of window and spectral leakage from sharp spectral features is well known in signal processing; see, e.g.,~\cite{Harris1978}.
In gravitational-wave detectors, there are a number of such features commonly referred to as ``lines''~\cite{Covas2018} or the low-frequency seismic wall.
The leading-order correction from the off-diagonal components of the covariance matrix is clearly seen using the following metric:
\begin{equation}
    \Delta_{i} \equiv \max_{i\neq j} \left|\frac{{C_{ij}}}{S_{j}}\right|.
\label{eq:contamination}
\end{equation}
We consider two regimes that determine the limiting behaviors of $\Delta_{i}$: (i) where the power spectral density is slowly varying (locally white noise) and (ii) near a large spectral feature.

We first consider the case of white noise, i.e., $\mathcal{C_{\mu\nu}} = \mathcal{S} \delta_{\mu\nu}$.
We write the finite-duration covariance matrix
\begin{equation}
    C_{ij} = \mathcal{S}\, \tilde{W}_{i\mu} \tilde{W}^{*}_{j\mu}.
\end{equation}
and
\begin{equation}
    \Delta_{i} = \max_{i\neq j} \frac{\left| \tilde{W}_{i\mu} \tilde{W}^{*}_{j\mu} \right|}{\left| \tilde{W}_{i\mu} \tilde{W}^{*}_{i\mu} \right|} = \max_{i \neq j} \frac{\left| \sum_{\mu} \tilde{w}_{i - \mu} \tilde{w}^{*}_{j - \mu} \right|}{\left| \sum_{\mu} \tilde{w}_{i - \mu} \tilde{w}^{*}_{i - \mu} \right|}.
\end{equation}
We note that the quantity $\tilde{W}_{i\mu} \tilde{W}^{*}_{j\mu}$ is real and its amplitude monotonically decreases as $|i - j|$ increases.
The contamination is therefore maximized when $i$ and $j$ are neighboring finite-duration frequency bins (we denote this as $j = i_{\pm}$ and emphasize that $i_{\pm} \neq i \pm 1$ due to the omitted interstitial frequencies in the finite-duration analysis).
We find
\begin{equation}\label{eq:leakage-white}
    \Delta_{i} = \frac{\left| \sum_{\mu} \tilde{w}_{i - \mu} \tilde{w}^{*}_{i_{\pm} - \mu} \right|}{\left| \sum_{\mu} \tilde{w}_{i - \mu} \tilde{w}^{*}_{i - \mu} \right|}.
\end{equation}
The variable $\Delta_i$ is a monotonically increasing function of the Tukey parameter $\alpha$, with $\Delta_{i}^{\alpha=0} = 0$ for a rectangular window and $\Delta_{i}^{\alpha=1} = 2/3$ for a Hann window (see Appendix~\ref{app:window-overlaps} for a derivation).
While the power spectrum of gravitational-wave detectors is not white, it is slowly varying away from the large spectral features and so we expect this approximation to hold throughout much of the observing band.

The other limiting case we can analytically describe is the behavior near a spectral line with relative amplitude $L$ at frequency $f_{\mu_l}$ with an otherwise white spectrum.
In this case we can approximate
\begin{equation}
    \mathcal{C}_{\mu\nu} = \begin{cases}
        \mathcal{S} & \mu = \nu \neq \mu_{l} \\
        L\, \mathcal{S} & \mu = \nu = \mu_{l} \\
        0 & \mu \neq \nu
    \end{cases}.
\end{equation}
We can now write the row corresponding to the line in the finite-duration covariance matrix:
\begin{align}
    C_{i\mu_{l}} &= \tilde{W}_{i\mu} \tilde{W}^{*}_{\mu_{l}\nu} \mathcal{C}_{\mu\nu} \\
    &= \mathcal{S} \begin{cases}
        \tilde{w}_{i - \mu} \tilde{w}^{*}_{\mu_{l} - \mu} & \mu \neq \mu_{l} \\
        L \tilde{w}_{i - \mu_{l}} \tilde{w}^{*}_{0} & \mu = \mu_{l}
    \end{cases} \\
    &= \mathcal{S} \left(\sum_{\mu\neq\mu_{l}} \tilde{w}_{i - \mu} \tilde{w}^{*}_{\mu_{l} - \mu} + L \tilde{w}_{i - \mu_{l}} \tilde{w}^{*}_{0} \right). \\
    &\approx L\, \mathcal{S} \tilde{w}_{i - \mu_{l}} \tilde{w}^{*}_{0}
\end{align}
In the last line we assume $L \tilde{w}_{i - \mu_{l}} \gg 1$ and so the contamination will fall off with the same spectral shape as the window function:
\begin{equation}
    \Delta_{i} \approx L\, \left| \tilde{w}_{i - \mu_{l}} \tilde{w}^{*}_{0} \right| \propto | \tilde{w}_{i - \mu_{l}} |.
\end{equation}
We emphasize at this stage that $\mu_{l}$ is not necessarily (and in fact almost guaranteed to not be) contained in the set of roman indices; i.e., the line is not exactly a delta function at one of the $1 / T$ Hz spaced frequency bins.
If the line were located at one of these frequencies, a rectangular window would have zero spectral leakage and be the optimal choice.
However, when this is not the case, the rectangular window maximizes the contamination from lines.
In practice, sharp spectral features in interferometer power spectra have finite width and therefore a rectangular window will never generically avoid leakage from lines.

In Fig.~\ref{fig:max-off-diagonal}, we show this quantity for the \ac{PSD}s used in our analysis of the gravitational-wave signal, GW170814~\cite{GW170814} (Sec.~\ref{sec:170814}), which was observed in 2017 by the Advanced LIGO~\cite{aLIGO} and Virgo~\cite{aVirgo} observatories.
For the two LIGO observatories~\cite{aLIGO}, the magnitude of the off-diagonal terms is approximately constant throughout the observing band with exceptions for the known lines.
The ``violin modes'' for the Livingston interferometer ($\sim \unit[500]{Hz}$) are significantly larger than for the Hanford interferometer and the contamination near the lines is larger and more broadband.
Given this behavior, one might think that we can neglect the impact of the off-diagonal terms if we remove from the analysis the frequency bins in the neighborhood of the lines.
However, for the Virgo observatory~\cite{aVirgo}, we see that the average correction is much larger across most of the band and is frequently $> 20\%$.
This can be attributed to the Virgo \ac{PSD} being less smoothly varying.
A cut based on frequencies with unacceptably large contamination would remove most of the observing band.
However, data analysis with the finite-duration covariance matrix can be used to take into account covariance in Virgo noise.

\subsection{Regularized inversion}\label{sec:inversion}

\begin{figure}
    \centering
    \includegraphics[width=\linewidth]{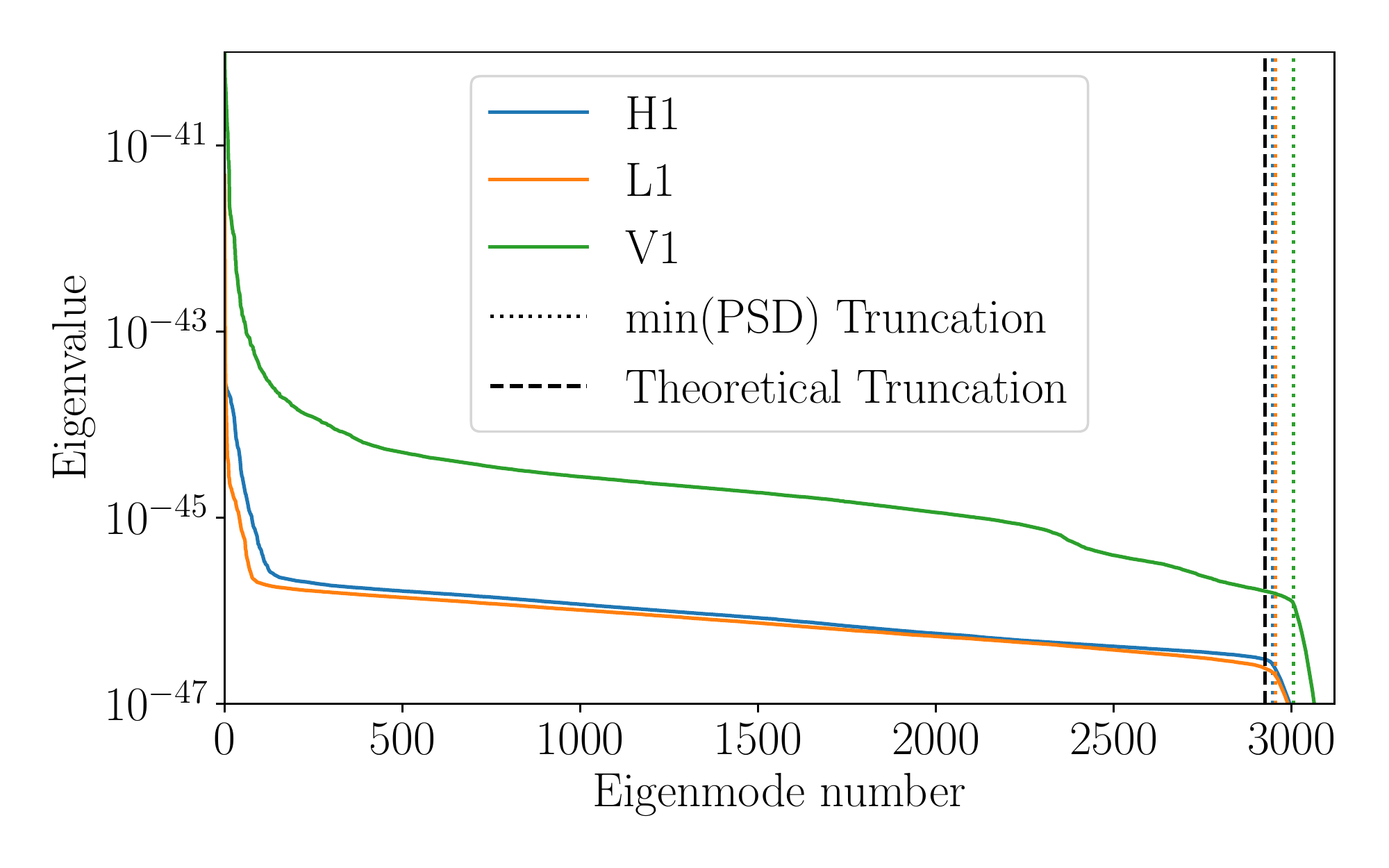}
    \caption{
    Eigenvalues of an estimated noise power covariance matrix for data close to GW170814 for a Tukey window with $\alpha=0.1$.
    The large eigenvalues correspond to spectral lines and the low-frequency seismic wall; the slowly varying region corresponds to the smoothly varying observing band.
    The dashed vertical line denotes the points after which we discard the eigenmodes as determined by the window information loss.
    The dotted lines indicate the point at which the eigenvalue drops below the minimum of the finite-duration \ac{PSD}.
    Both of these well-approximate the turnover after which the eigenvalues rapidly decline due to information loss from windowing.
    The difference is most pronounced for the Virgo data, which is consistent with the increased contamination between frequency bins (see Figure~\ref{fig:delta}).
    }
    \label{fig:eigenvalues}
\end{figure}

\begin{figure}
    \centering
    \includegraphics[width=\linewidth]{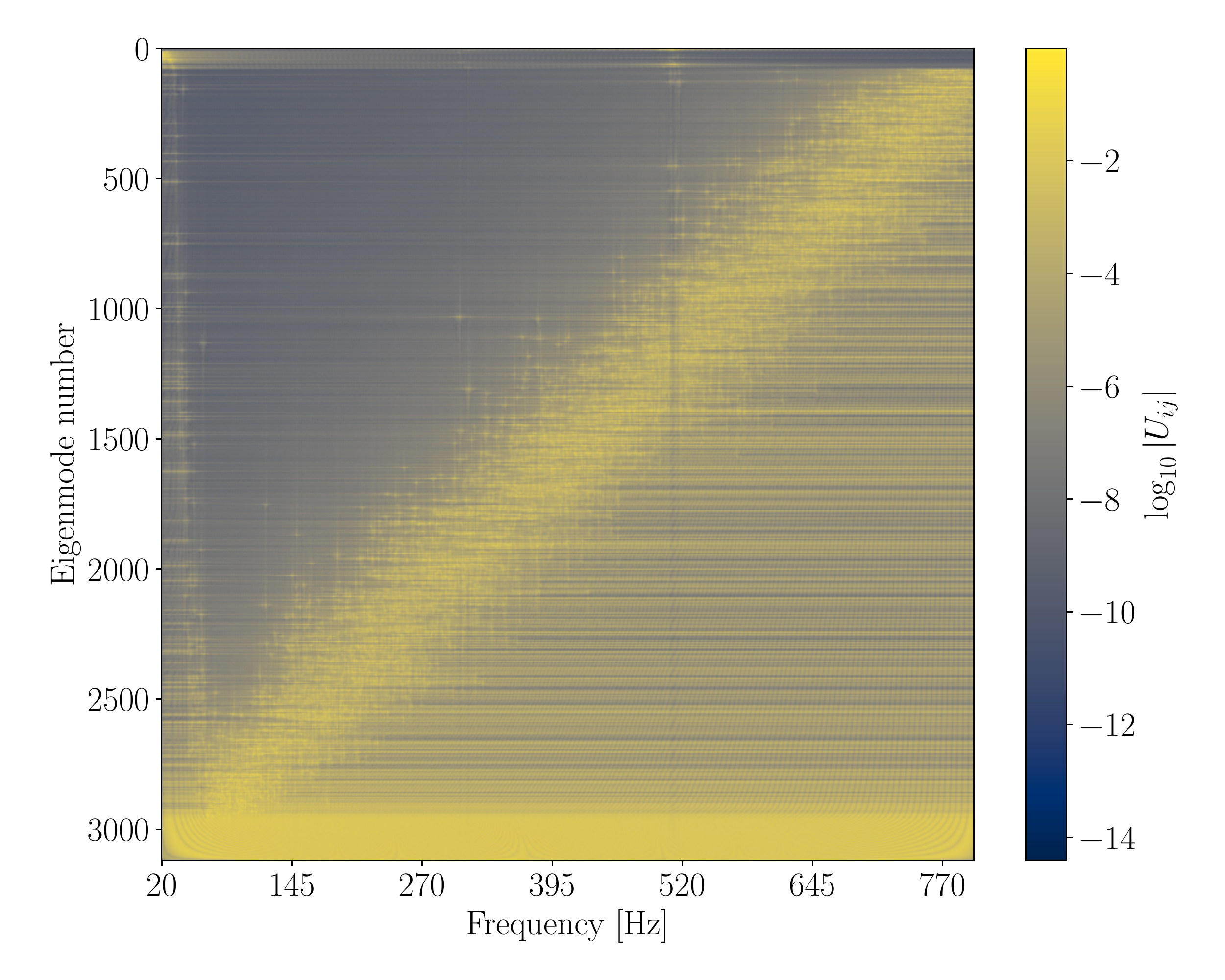}
    \caption{
    Eigenmodes for the covariance matrix estimated from data from the LIGO Livingston interferometer close to GW170814.
    This matrix encodes the correlations between physical frequencies.
    The horizontal axis corresponds to the physical frequencies, while the vertical axis is the order of decreasing eigenvalue.
    The bottom-most eigenmodes are the ones that are discarded.
    These eigenmodes are associated with very broadband frequency content.
    }
    \label{fig:eigenmodes}
\end{figure}

Having characterized the covariance between frequency bins due to windows, we turn to the inversion of the covariance matrix required to evaluate the likelihood function.
Since tapered window functions go to zero at the edges by construction, the covariance matrix is not invertible~\footnote{The covariance matrix is non-invertible as the window application is not reversible. There is no way to recover the value of the time series for points at which the data have been zeroed, e.g., the end points.)}.
To deal with this issue, we perform a regularized inversion using a \ac{SVD} and discard the smallest eigenvalues.
The \ac{SVD} of a Hermitian matrix can be written as
\begin{align}
    C_{ij} = U_{ik} \Lambda_{kl} U^{-1}_{lj}.
\label{eq:svd}
\end{align}
Here, $\Lambda_{kl} = \delta_{kl} \lambda_{k}$ (no summation) is a diagonal matrix with the eigenvalues $\lambda_{k}$ on the leading diagonal.
Regularization simply involves removing eigenmodes corresponding to small eigenvalues.
In practice, this is done by setting the eigenvalue to $\infty$:
\begin{equation}
    \bar{\lambda}_{i} = \begin{cases}
    \lambda_{i} \qquad i \leq \epsilon N \\
    \infty \qquad i > \epsilon N
    \end{cases}
\label{eq:regularize}
\end{equation}
where a fraction $\epsilon$ of the $N$ eigenmodes are retained.
The regularized matrix and its inverse are
\begin{align}
    \bar{C}_{ij} = & U_{ik} \bar{\Lambda}_{kl} U^{-1}_{lj} \label{eq:regularized-covariance} \\
    \bar{C}^{-1}_{ij} = & U_{ik} \bar{\Lambda}^{-1}_{kl} U^{-1}_{lj}.
    \label{eq:regularized-inverse-covariance}
\end{align}

In Fig.~\ref{fig:eigenvalues}, we show the eigenvalues in decreasing order for the covariance matrix estimated at the time of GW170814 and the corresponding eigenmode spectrum is shown in Fig.~\ref{fig:eigenmodes}.
We identify three regimes in the eigenvalue spectrum:
\begin{enumerate}
    \item Large eigenvalues with a steep slope at low eigenmode number. These predominantly correspond to frequencies where $C_{ij}$ is large, e.g., near spectral lines and at low frequencies.
    \item A slowly varying region encompassing the majority of the eigenvalues. This corresponds to the remainder of the frequencies where the PSD is smoothly varying.
    \item A rapid drop to the smallest eigenvalues. This is a characteristic feature of ill-conditioned matrices and is the region we should remove when regularizing.
\end{enumerate}

We consider two methods to determine how many eigenvalues to discard.
In the first method we consider the power loss from the window function.
The amount of information lost by the window is related to the time-averaged square of the window function.
The effective number of independent time samples is
\begin{align}
    N_\text{eff} = N \overline{w^2} 
    = N\int dt \, w^2(t) \approx \sum_{i=0}^{N - 1} w^{2}_{i}.
\end{align}
We choose the fraction of eigenvalues to retain based on the window function:
\begin{equation}
    \epsilon = \frac{N_{\text{eff}}}{N} = \overline{w^2}.
\end{equation}
The vertical dashed line is at $N_\text{eff}$ and shows the number of modes we omit to account for the information loss due to the tapered window; we note that this matches well the transition to the badly behaved modes.
For the second method we set a threshold value corresponding to the minimum of the power spectral density over the analyzed frequency band $\lambda_{T} = \min_{\mu}(\mathcal{S}_{\mu})$.
We show this threshold in the dotted line in Figure~\ref{fig:eigenvalues}.
The two threshold values agree well for small Tukey $\alpha$ although the former method systematically removes more eigenmodes.
We find that the choice of regularization scheme has a negligible impact on the results of our analysis.

\subsection{Finite-duration likelihood}
We can now write our final result, the regularized likelihood with a finite-duration covariance matrix:
\begin{equation}
    \bar{\mathcal{L}}(\tilde{d} | \theta, \bar{C}) = \frac{2}{T\det{\bar{C}}} \exp \left[ - \frac{2}{T} \left< \tilde{d} - \tilde{h}, \tilde{d} - \tilde{h} \right>_{\bar{C}} \right],
\label{eq:finite-likelihood}
\end{equation}
where $\det \bar{C}$ is the determinant of the finite-duration noise covariance matrix, $\tilde{h}$ is a template for the signal, and the inner product is defined as
\begin{equation}\label{eq:inner-product}
    \left< \tilde{x}, \tilde{x} \right>_{\bar{C}} = \tilde{x}_i \bar{C}_{ij}^{-1} \tilde{x}^{*}_j.
\end{equation}
As the exponent in the likelihood can still be written as weighted inner products between data and template, we can analytically marginalize over extrinsic parameters in the same way as for the diagonal likelihood~\cite{Thrane2019}.

\subsection{Relation to the Karhunen-Lo\`{e}ve transform}

The Karhunen-Lo\`{e}ve theorem states that for any stochastic process there exists a basis in which the noise covariance matrix is diagonal, and the transformation into this basis is referred to as the \ac{KLT}.
For a colored stationary Gaussian process that is periodic with period $T$, this basis is the discrete Fourier transform with spacing $1 / T$ Hz.
The Whittle likelihood approximation specifically assumes that this basis also diagonalizes the covariance matrix for a subset of a longer Gaussian process that is not periodic with period $T$.
As we have demonstrated, the covariance matrix in the Fourier basis is not diagonal in this case.

We note that the \ac{KLT} is closely related to the \ac{SVD} and therefore identify that the basis for the \ac{KLT} of a finite subset of a longer Gaussian process is the basis found in Section~\ref{sec:inversion}.
This provides a second, equivalent, interpretation of the inner product in Eq.~\ref{eq:inner-product}:
\begin{align}
    \left< \tilde{x}, \tilde{x} \right>_{\bar{C}}
    &= \tilde{x}_i U_{ik} \bar{\Lambda}^{-1}_{kl} U^{-1}_{lj}
    = \sum_{i} \frac{|\bar{x}_i|^2}{\bar{\lambda}_{i}},
\end{align}
where we have defined $\bar{x}_i \equiv \tilde{x}_{i} U_{ik}$.
The likelihood is now
\begin{equation}
    \ln \mathcal{L} = - \sum_{i} \left[ \frac{|\bar{d}_i - \bar{h}_i|^2}{\lambda_{i}} + \ln\left( \lambda_{i} \right) \right] + {\rm const.}
\end{equation}
We identify that this has the usual form of the likelihood except that all of the quantities are described in the eigenbasis of the \ac{KLT}, rather than the Fourier basis.

\section{Demonstration}\label{sec:demonstration}

To demonstrate our formalism we perform two tests.
First, we analyze three binary black hole mergers to demonstrate that the effect of the off-diagonal corrections is small but noticeable for confidently detected signals.
Second, we demonstrate that, although this effect produces a minor correction to modest-\ac{SNR} events, neglecting the impact of off-diagonal terms in the noise covariance matrix biases precision estimates, such as evidence calculations required for searches for a population of weak, sub-threshold signals as in~\cite{Smith2018}.

\subsection{Single events}\label{sec:170814}

\begin{figure}
    \centering
    \includegraphics[width=\linewidth]{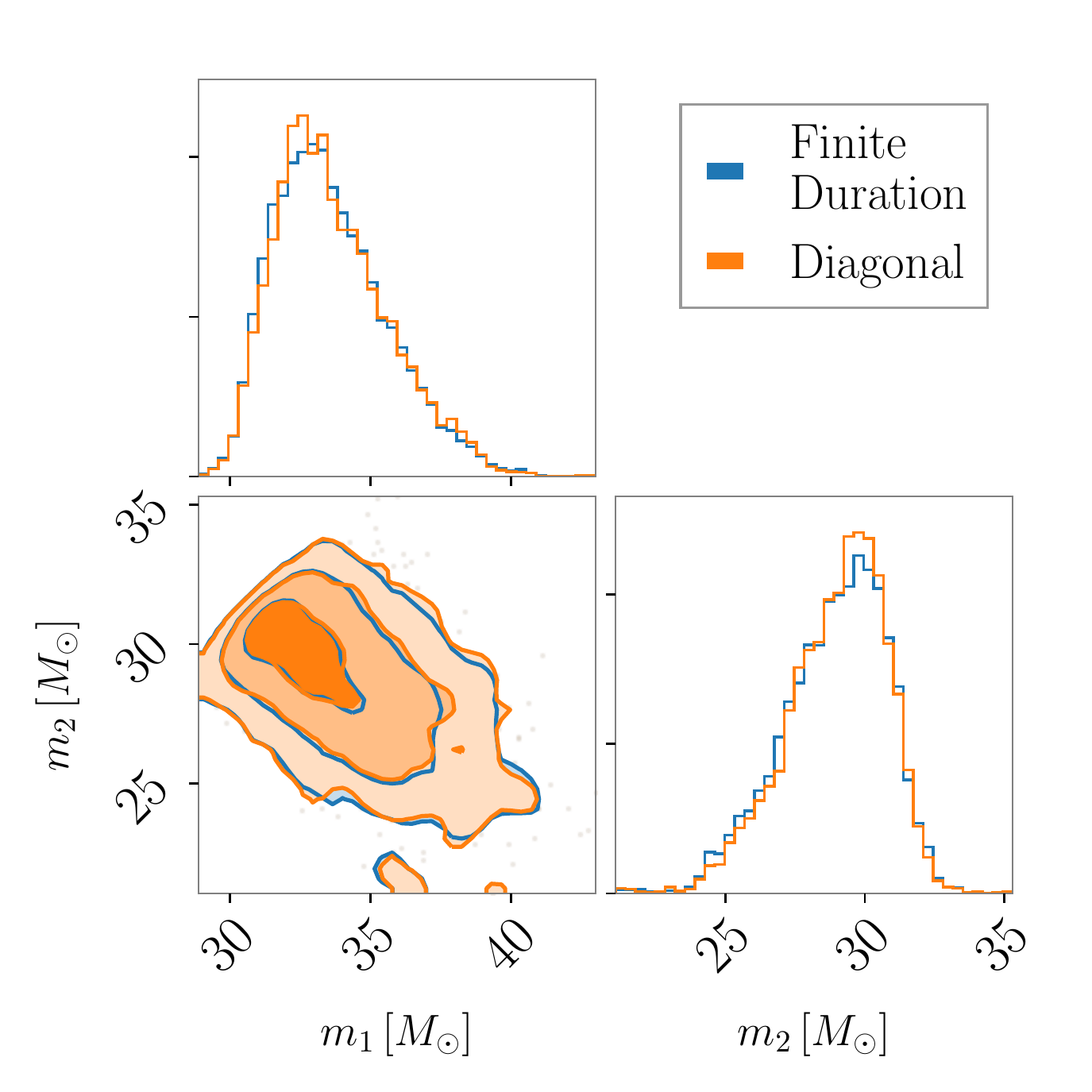}
    \caption{Intrinsic parameters for the binary black hole merger GW150914~\cite{GW150914} using our new finite-duration likelihood (blue) and the diagonal likelihood (orange).
    The primary and secondary mass ($m_1$, $m_2$) refer, respectively, to the more-massive and less-massive component masses.
    Including covariance between neighboring bins has no observable impact on the inferred posterior.
    }
    \label{fig:150914-intrinsic}
\end{figure}

\begin{figure}
    \centering
    \includegraphics[width=\linewidth]{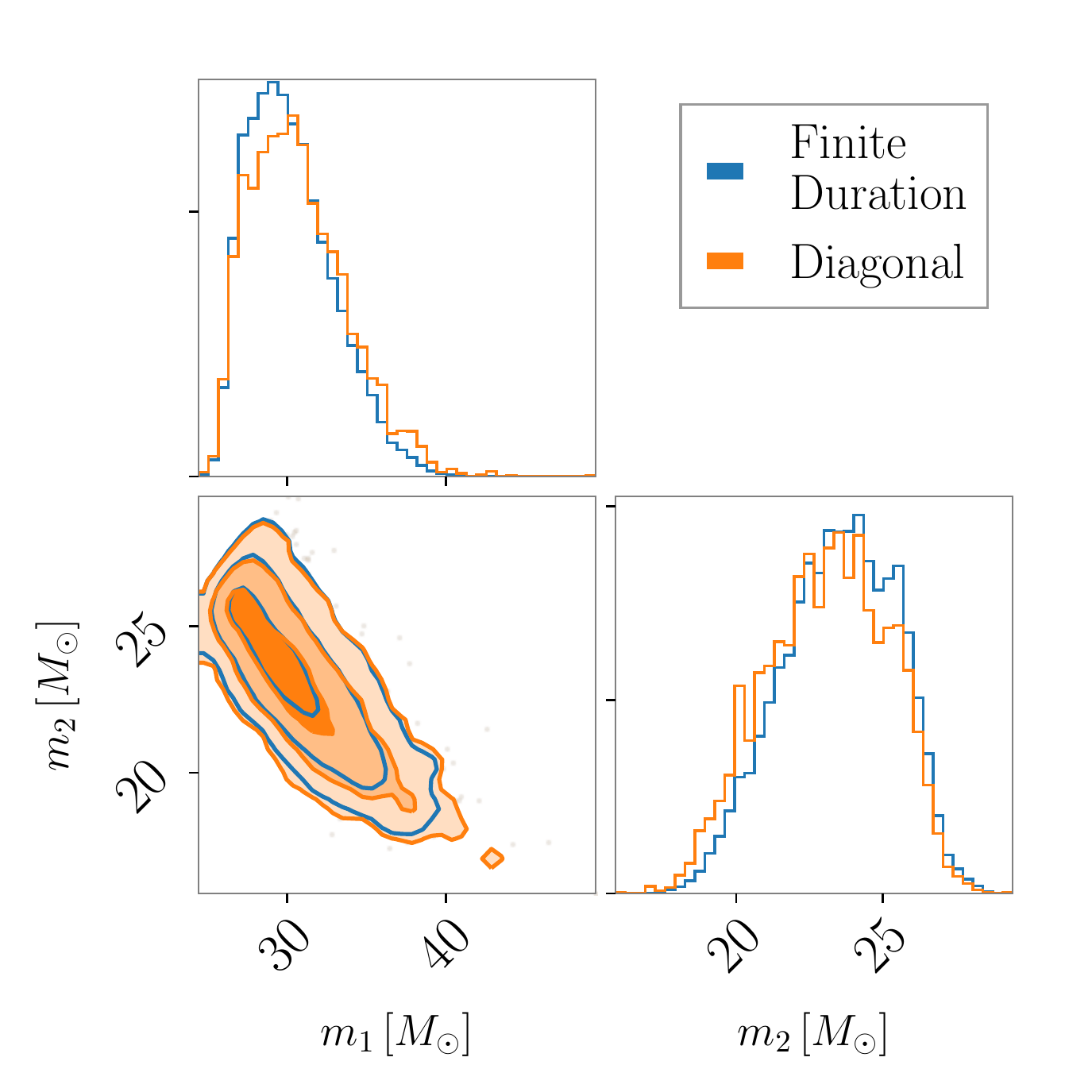}
    \caption{Intrinsic parameters for the binary black hole merger GW170814~\cite{GW170814} using our new finite-duration likelihood (blue) and the diagonal likelihood (orange).
    The primary and secondary mass ($m_1$, $m_2$) refer, respectively, to the more-massive and less-massive component masses.
    Including covariance between neighboring frequency bins slightly shifts the inferred posterior distribution for the mass ratio.
    }
    \label{fig:170814-intrinsic}
\end{figure}

\begin{figure}
    \centering
    \includegraphics[width=\linewidth]{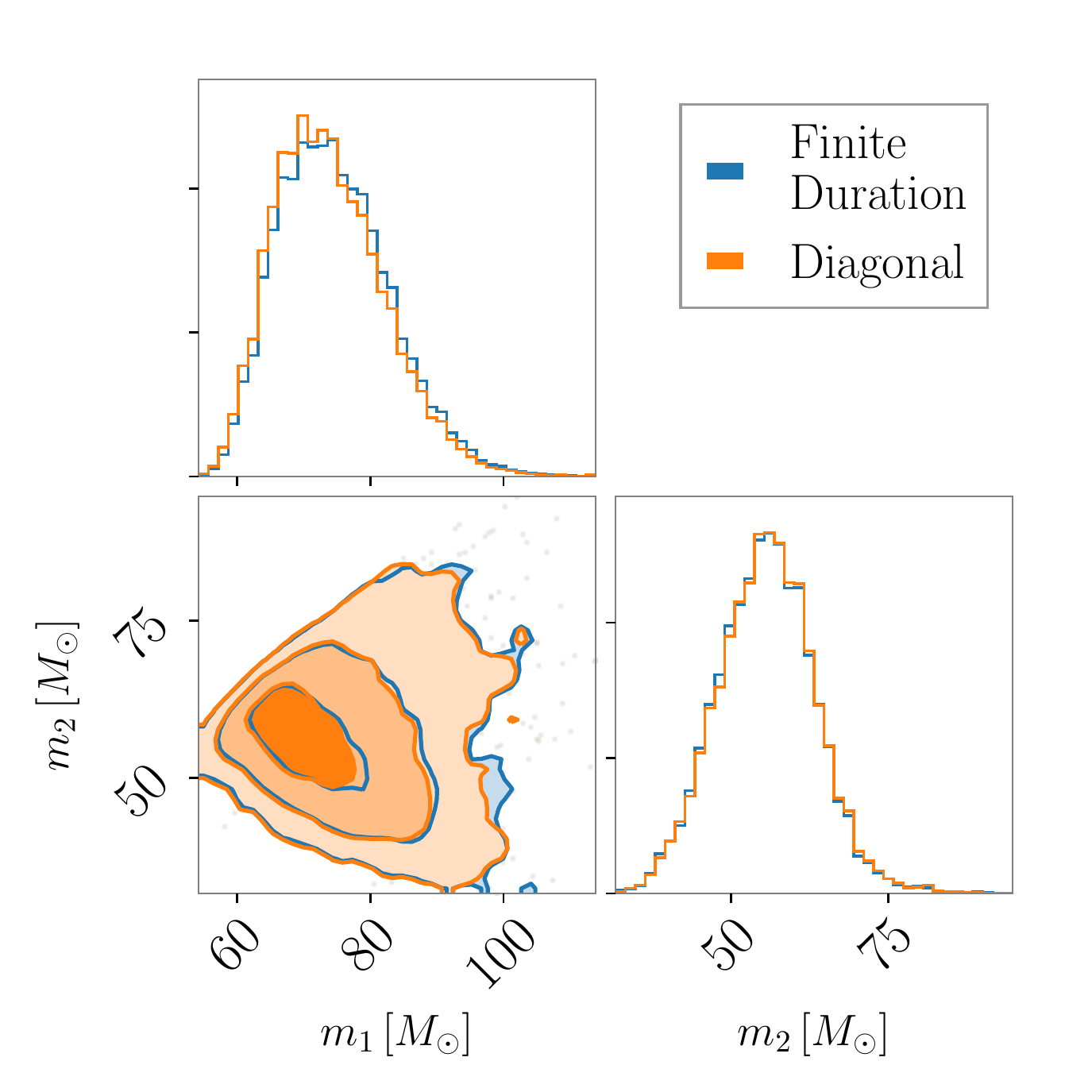}
    \caption{Intrinsic parameters for the binary black hole merger GW190521~\cite{GW190521} using our new finite-duration likelihood (blue) and the diagonal likelihood (orange).
    The primary and secondary mass ($m_1$, $m_2$) refer, respectively, to the more-massive and less-massive component masses.
    Including covariance between neighboring bins has no observable impact on the inferred posterior.
    }
    \label{fig:190521-intrinsic}
\end{figure}

We compare the posterior distributions obtained using both the conventional and finite-duration covariance matrices for three of the observed binary black hole mergers GW150914~\cite{GW150914}, GW170814~\cite{GW170814}, and GW190521~\cite{GW190521}.
We choose these as they are relatively high-mass systems (with detector frame primary and secondary masses of $m_1\approx m_2 \approx 30-40 M_{\odot}$ for GW150914 and GW170814 and $m_1 \approx m_2 \approx 150 M_{\odot}$ for GW190521) for which we expect the tapered window to have a comparatively large impact on the data as they use a relatively short stretch of analysis data.
They also span a large range in time, with one event from each of the first three observing runs of the advanced detector network, leading to significantly different PSDs.

For each event, we analyze $\unit[4]{s}$ of data centered at the trigger time for the event and estimate the \ac{PSD} using a $\unit[512]{s}$ stretch of data ending $\unit[2]{s}$ before the trigger.
We apply a bandpass filter between $16$ and $\unit[1024]{Hz}$ and resample the data to a new Nyquist frequency of $\unit[2048]{Hz}$ using \texttt{GWpy}~\cite{gwpy} to mitigate spectral leakage from low and high frequencies.
We use a Tukey window with $\alpha=0.1$ for the analysis segment.
The details of the covariance matrix calculation are described in Algorithm~\ref{algo:psd-calculation}.
We employ the \texttt{IMRPhenomXPHM} waveform approximant~\cite{Pratten2020, GarciaQuiros2020, Pratten2021} in the frequency range $\unit[20 - 800]{Hz}$; for GW190521 we set the upper frequency limit as $\unit[300]{Hz}$.
We neglect the impact of calibration uncertainty or uncertainty in our estimate of the \ac{PSD}.
For GW150914, we analyze data from the two LIGO interferometers, for the other two events, we analyze data from the two LIGO interferometers and Advanced Virgo.

We show the posterior distribution for two of the intrinsic binary parameters when assuming $C_{ij}$ is diagonal (blue) and using the full covariance matrix (orange) for GW150194, GW170814, and GW190521 in Figs.~\ref{fig:150914-intrinsic},~\ref{fig:170814-intrinsic}, and~\ref{fig:190521-intrinsic} respectively.
We assume the same prior for both analyses.
The primary and secondary mass refer, respectively, to the more-massive and less-massive black-hole masses.
The largest difference we see is in the component masses for GW170814, primarily driven by a change in the inferred mass ratio.
There is no visible difference between the posteriors for the other events.
The change in the posterior distributions is at a similar level to the changes due to marginalizing over uncertainty in the detector calibration~\cite{Payne2020, Vitale2021} or \ac{PSD} estimate~\cite{Talbot2020, Biscoveanu2020,Chatziioannou2019}, but less than the difference due to using different \ac{PSD} estimation methods~\cite{Talbot2020}.
Errors of this magnitude become important when we combine many events together for population studies and/or precision tests of general relativity (see, e.g., \cite{Purrer2020,Moore2021}).

\subsection{Combining data segments}\label{sec:tbs}
By combining large numbers of time-series data segments it is sometimes possible to extract weak signals not visible in individual segments, for example, to measure the population of gravitational waves from unresolved compact binaries~\cite{Smith2018,Gaebel2019,Smith2020, Banagiri2020b} and to detect gravitational-wave memory~\cite{Lasky2016}.
Combining data segments can also have the effect of magnifying systematic errors that are small enough to ignore when considering just a single segment in isolation.
For example, failing to take into account uncertainty in estimates of the noise \ac{PSD} leads to low-level excess power, which can be mistaken for a population of sub-threshold gravitational-wave signals~\cite{Biscoveanu2020,Talbot2020}.

Here, we show that the correlations between neighboring frequency bins induced by all windows must be taken into account to avoid systematic error in studies that rely on precision measurements combining many segments.
We illustrate this point using simulated data to carry out a mock search for a population of sub-threshold simulated signals in simulated Gaussian noise with a known \ac{PSD}.
We employ the formalism from~\cite{Smith2018} to estimate the fraction of $M=160000$ data segments of which 15000 contain a simulated signal.

Our likelihood is a mixture model, which allows for each segment to consist of either signal ${\cal S}$ or noise ${\cal N}$:
\begin{align}
    {\cal L}(\{d\}|\xi) = \prod_i^M \Big[\xi{\cal L}(d_i|{\cal S}) + (1-\xi){\cal} {\cal L}(d_i|{\cal N}) \Big].
\end{align}
Here, ${\cal L}(d_i|{\cal S})$ is the likelihood of data segment $i$ given the signal hypothesis while ${\cal L}(d_i|{\cal N})$ is the likelihood given the noise hypothesis and the parameter $\xi$ is the fraction of segments that contain a signal.
We simulate data in $\unit[128]{s}$ chunks and break up the data into $\unit[4]{s}$ segments.
We compute the finite-duration \ac{PSD} matrix using the known \ac{PSD} used to simulate the data.
The known \ac{PSD} is as estimated for the LIGO Livingston detector in our analysis of GW170814.
We do not re-estimate the \ac{PSD} from the simulated data in order to avoid uncertainty intrinsic to the \ac{PSD} estimation method.

\begin{figure}
    \centering
    \includegraphics[width=\linewidth]{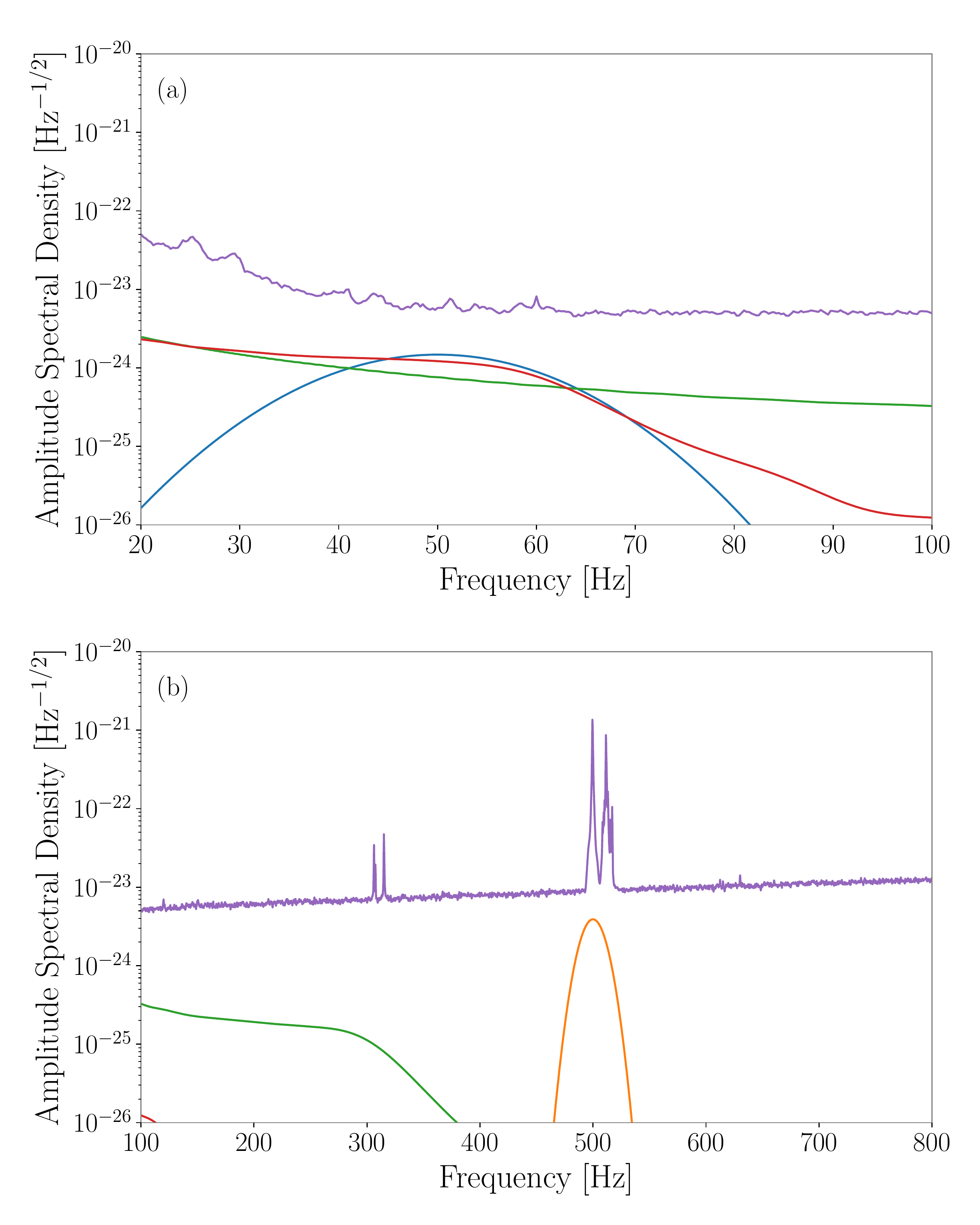}
    \caption{
        The signals considered for our population test.
        The four signals we consider are: a Gaussian burst centered at 50 Hz and standard deviation $\unit[10]{Hz}$ (blue), a Gaussian burst centered at $\unit[500]{Hz}$ and standard deviation 10 Hz (orange), a $60 M_{\odot}$ binary black hole merger (green), and a $300 M_{\odot}$ binary black hole merger (red).
        In purple, we show the finite-duration amplitude spectral density for the LIGO Livingston interferometer at the time of binary black hole merger GW170814.
        The Gaussian bursts isolate the impact of specific spectral features, e.g., the large lines around $\unit[500]{Hz}$.
        The binary black hole mergers are broadband and accumulate most of their signal-to-noise ratio at low frequencies, near the sharp rise due to seismic noise.
    }
    \label{fig:tbs_signals}
\end{figure}

\begin{figure}
\centering
\includegraphics[width=\linewidth]{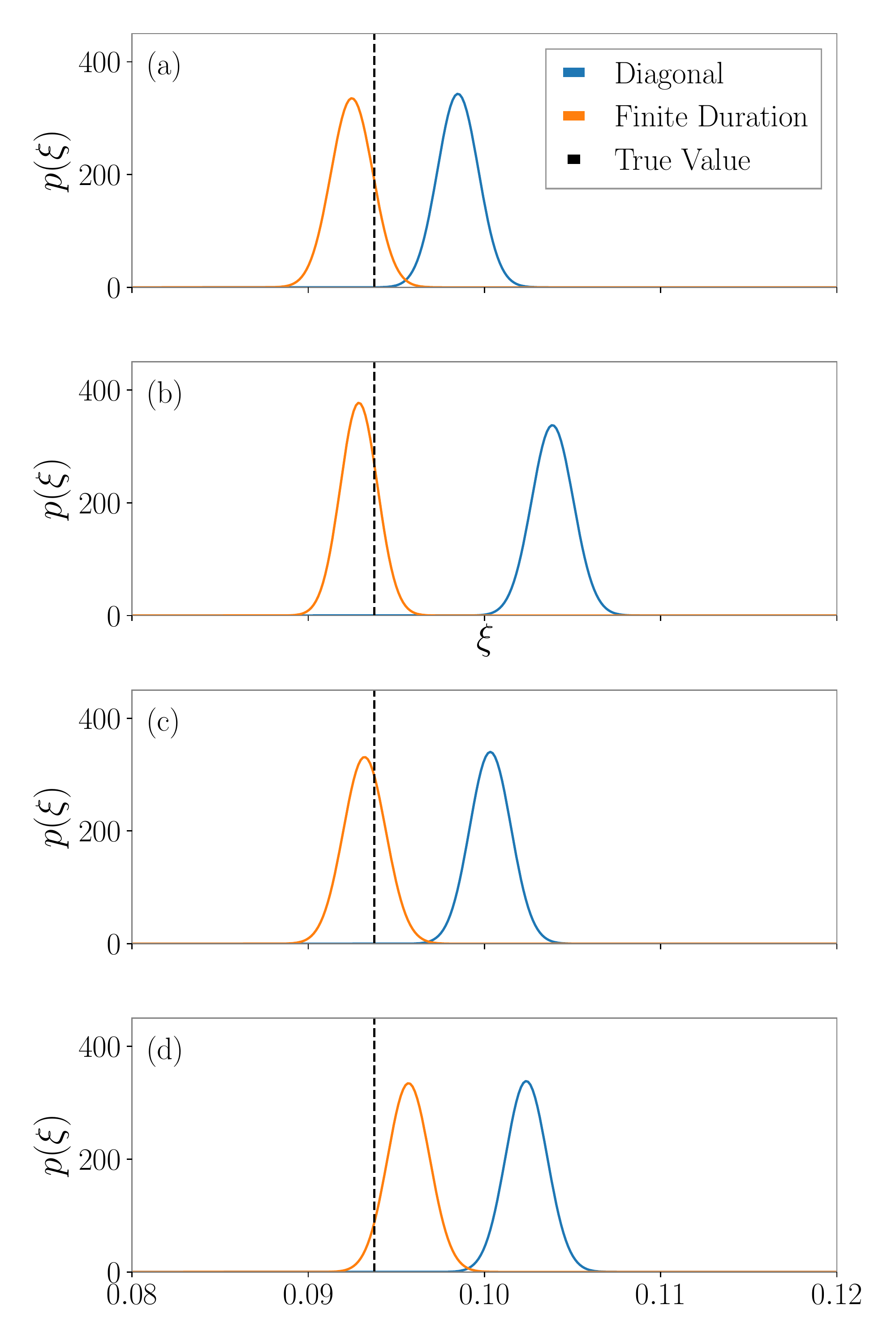}
\caption{
The posterior distribution for the fraction of segments containing a signal in our toy example (Section~\ref{sec:tbs}).
We analyze $\sim 7.5$ days of simulated Gaussian noise divided into $4s$ segments ($1.6\times10^5$ segments), and 15000 of these segments contain four different signals: (i) a Gaussian pulse with standard deviation $\unit[10]{Hz}$ and central frequency $\unit[50]{Hz}$ (ii) a Gaussian pulse with central frequency $\unit[500]{Hz}$, (iii) an equal-mass binary black hole merger signal with total mass $60 M_{\odot}$ and (iv) an equal-mass binary black hole merger signal $300 M_{\odot}$.
The amplitude of the signals and the \ac{PSD} are shown in Figure~\ref{fig:tbs_signals}.
In blue we show the posterior distribution obtained using the standard likelihood that ignores correlations between neighboring frequency bins induced by the window function.
In orange, we show the posterior distribution using a likelihood that accounts for these correlations.
We see that the diagonal method gives a biased result when the signal is centered at $\unit[500]{Hz}$ and when analyzing simulated binary black hole systems.
}
\label{fig:tbs}
\end{figure}

For the simple example considered here we assume that the signal in each segment containing a signal is the same~\footnote{We emphasize that this method trivially extends to the more generic case of an unknown population of signals, however, we make this assumption to isolate the impact of the covariance matrix.}.
We consider four choices of signal: (i) a Gaussian burst centered at $\unit[50]{Hz}$ and standard deviation $\unit[10]{Hz}$, (ii) a Gaussian burst centered at $\unit[500]{Hz}$ and standard deviation $\unit[10]{Hz}$, (iii) a $60 M_{\odot}$ total mass binary black hole merger waveform, and (iv) a $300 M_{\odot}$ total mass binary black hole merger waveform.
The first two are two well-localized signals in frequency with random per-frequency phases.
The lower-frequency burst does not significantly overlap with large spectral features, while the higher-frequency burst overlaps with the largest spectral lines.

The second two are representative of the gravitational-wave signals observed so far and considered in the search proposed in~\cite{Smith2018}.
These signals are relatively broadband in frequency with well-defined frequency evolution.
The $60 M_\odot$ waveform is chosen to be representative of the most commonly observed systems and the $300 M_\odot$ waveform is chosen based on the largest observed system~\footnote{The masses presented here are ``detector-frame'' quantities which are larger than the source-frame masses due to cosmological redshift.}.
As in the previous section, for the binary black hole waveforms we use the \texttt{IMRPhenomXPHM} waveforms.
In Figure~\ref{fig:tbs_signals}, we show the amplitude spectral density of each of the signals along with the diagonal of the finite-duration covariance matrix ($S_{i}$).

For each iteration, we calculate the posterior for $\xi$ two ways, once using the standard diagonal likelihood and once using the \new{} likelihood.
The results are shown in Fig.~\ref{fig:tbs} for a Tukey window with $\alpha=0.1$.
For all considered signals, the diagonal method produces a visibly biased result.
When the Gaussian signal is close to large spectral lines (b) the bias is most significant.
The bias for the binary black hole signals (c-d) is larger than for the low-frequency Gaussian (a).
We attribute this to two effects.
The binary black hole signals accumulate significant \ac{SNR} at low frequencies where the noise is dominated by leakage from seismic noise.
This is somewhat mitigated by the application of a high-pass filter to suppress content below $\unit[16]{Hz}$.
They also have a characteristic phase evolution, which contributes significant resolving power to the likelihood function.
This means that the phase coherence between the neighboring frequencies is important to the evaluated likelihoods.

\section{Discussion}\label{sec:conclusions}
As gravitational-wave astronomy matures, the growing catalog of events enables exciting new science.
However, as we probe increasingly higher \ac{SNR}s, and as we combine larger ensembles of data segments, our analyses are increasingly susceptible to systematic error from approximations in our models and data analysis.
Many sources of systematic error in our modeling have been considered in recent years including calibration uncertainty~\cite{Payne2020, Vitale2021}, waveform systematics~\cite{Ashton2020, Purrer2020, Jan2020, Estelles2021}, and noise estimation~\cite{Talbot2020, Biscoveanu2020, Chatziioannou2019}.
In this paper, we examine how correlations between frequency bins are inevitably introduced by windowing in time-domain analysis and find corrections at the same level as those due to calibration and \ac{PSD} uncertainty.
We show how these correlations can be modeled using a frequency-domain noise covariance matrix, thereby avoiding bias.
By performing a singular value decomposition, we identify that the basis that diagonalizes the covariance matrix is not the Fourier basis as usually assumed, but depends on both the \ac{PSD} and the choice of window function.

We demonstrate that, while the impact of the off-diagonal components in the noise covariance matrix is small for individually resolved events, they are important for precision estimates of the Bayesian evidence.
These precision estimates of the Bayesian evidence are crucial when attempting to use Bayesian evidence estimates as a detection statistic, e.g.,~\cite{Smith2018, bcr}.

A natural extension of the framework provided here is to incorporate marginalization over sources of systematic uncertainty.
Marginalization over waveform or detector calibration uncertainty can be trivially combined with this method.
Marginalizing over uncertainty in the \ac{PSD} estimate would require either modifying the \texttt{BayesLine} algorithm~\cite{Littenberg2015} to include modeling the full noise covariance matrix or an analytic method such as in~\cite{Rover2011, Banagiri2020a, Talbot2020}.
Even after considering all these forms of systematic uncertainty, we still need to develop methods to deal with the non-Gaussianity and non-stationarity of real data.
This is an active area of development~\cite{Tiwari2015, Pankow2018, Zackay2019, Cornish2020, Ormiston2020, Chatziioannou2021, Edy2021, Mogushi2021}; however, establishing a unified treatment is left to future studies.

While the analysis here has focused on analysis of short-duration transients, windows are also used in searches for longer duration gravitational-wave signals~\cite{Romano2017, Sieniawska2019}.
Typically, long-duration searches use much longer segment durations than those described here with Hann windows to mitigate leakage from lines.
As we showed in this paper, this may lead to correlations between neighboring frequency bins for these analyses.
In searches for the stochastic gravitational-wave background, a coarse-grained \ac{PSD} estimate is typically used (see Section~\ref{appendix:coarse-grain}) which may reduce the impact of these correlations.
Additionally, window functions are used to excise short duration non-Gaussian ``glitches'' from analysis~\cite{170817, Pankow2018, O3Stochastic, Gating} (typically this is referred to as ``gating'').
These gates have very short rise times (small Tukey $\alpha$) and so introduce significant contamination around spectral lines.

\section*{Acknowledgements}
We are grateful to Sharan Banagiri, Katerina Chatziiaonnou, Joe Romano, and Alan Weinstein for many fruitful discussions and the anonymous referee for a thoughtful and detailed review.
CT acknowledges the support of the National Science Foundation, and the LIGO Laboratory.
This work is supported through the Australian Research Council (ARC) Centre of Excellence CE170100004. SB is also supported by the Paul and Daisy Soros Fellowship for New Americans, the Australian-American Fulbright Commission, and the NSF Graduate Research Fellowship under Grant No. DGE-1122374.
This research has made use of data, software, and/or web tools obtained from the Gravitational Wave Open Science Center~\cite{Vallisneri2015, OpenData} (\href{https://www.gw-openscience.org}{https://www.gw-openscience.org}), a service of LIGO Laboratory, the LIGO Scientific Collaboration and the Virgo Collaboration. LIGO is funded by the U.S. National Science Foundation. Virgo is funded by the French Centre National de Recherche Scientifique (CNRS), the Italian Istituto Nazionale della Fisica Nucleare (INFN), and the Dutch Nikhef, with contributions by Polish and Hungarian institutes.
The authors are grateful for computational resources provided by the LIGO Lab and supported by National Science Foundation Grants PHY-0757058 and PHY-0823459.
This is document LIGO-P2100090.

\texttt{Cython} and \texttt{CUDA} implementations and \texttt{Python} wrappers of the \ac{PSD} matrix calculation are available at \href{https://github.com/ColmTalbot/psd-covariance-matrices}{github.com/ColmTalbot/psd-covariance-matrices}.
We also provide example scripts demonstrating how to produce the results in this paper and some supplementary results in the same location.

\begin{acronym}
\acro{PSD}[PSD]{power spectral density}
\acro{SNR}[SNR]{signal-to-noise ratio}
\acro{SVD}[SVD]{singular value decomposition}
\acro{KLT}[KLT]{Karhunen-Lo\`{e}ve transformation}
\end{acronym}

\appendix

\section{Connection to coarse-grained PSD estimation}\label{appendix:coarse-grain}

While time-averaging of short segments to estimate power spectral densities (e.g., Welch averaging) is common in gravitational-wave data analysis, an alternative ``coarse-graining'' method is used in some areas, especially searches for the stochastic gravitational-wave background; see, e.g.,~\cite{S6Isotropic}.
In this appendix, we demonstrate that coarse graining is a special case of the projection method we use in this paper.

The coarse-grained \ac{PSD} is defined for a frequency resolution $\delta f$ as
\begin{align}
    S_{i}
    &= \frac{1}{\delta f} \int^{f_i + \delta f / 2}_{f_i - \delta f / 2} df \mathcal{S}(f)
    \\
    &= \frac{1}{\delta f} \int^{\infty}_{-\infty} df \Pi(f_i - \delta f / 2, f_i + \delta f / 2) \mathcal{S}(f)
    \\
    &= \frac{1}{\alpha} \mathcal{S}_{\mu} |\tilde{w}_{\mu - \alpha i}|^2
    \quad \alpha \in \mathbb{Z} \\
\tilde{w}_{\mu} &= \begin{cases}
    1 &\quad -\frac{\alpha}{2} < \mu < \frac{\alpha}{2} \\
    0.5 &\quad |\mu| = \frac{\alpha}{2} \quad {\rm if}\,\, \frac{\alpha}{2} \in \mathbb{Z} \\
    0 &\quad \text{else}
    \end{cases}
\end{align}
In the second line, $\Pi$ is the unit boxcar function.
As for the window operator in this paper, we can express this as a circulant matrix with non zero entries only in a small region.
The corresponding time-domain window is the sinc function.
We note that the coarsened frequency-domain covariance matrix is diagonal by construction in this case.

\section{Estimating the infinite-duration \ac{PSD}}\label{app:infinite-duration}

Computing $C_{ij}$ using Equation~\ref{eq:exact-covariance} requires an expression for the infinite-duration covariance matrix $\mathcal{C_{\mu\nu}}$.
In this Appendix, we describe our method for estimating $\mathcal{C}_{\mu\nu}$ from the data.
This represents stages $1-4$ of Algorithm~\ref{algo:psd-calculation}.

We assume the data are stationary and Gaussian and therefore the only nonzero elements of $\mathcal{C}_{\mu\nu}$ are the leading diagonal $\mathcal{S}_{\mu}$.
We approximate this infinite-duration \ac{PSD} using segments longer than our final analysis segment.
In order to avoid long-term drift of $\mathcal{S}_{\mu}$ in real interferometer data, we restrict ourselves to $\unit[512]{s}$ of data.
We then subdivide this into non-overlapping segments with duration = $D$ and compute a median average \ac{PSD} using a Hann window as our representation of $\mathcal{S}_{\mu}$.

\begin{figure}
    \centering
    \includegraphics[width=\linewidth]{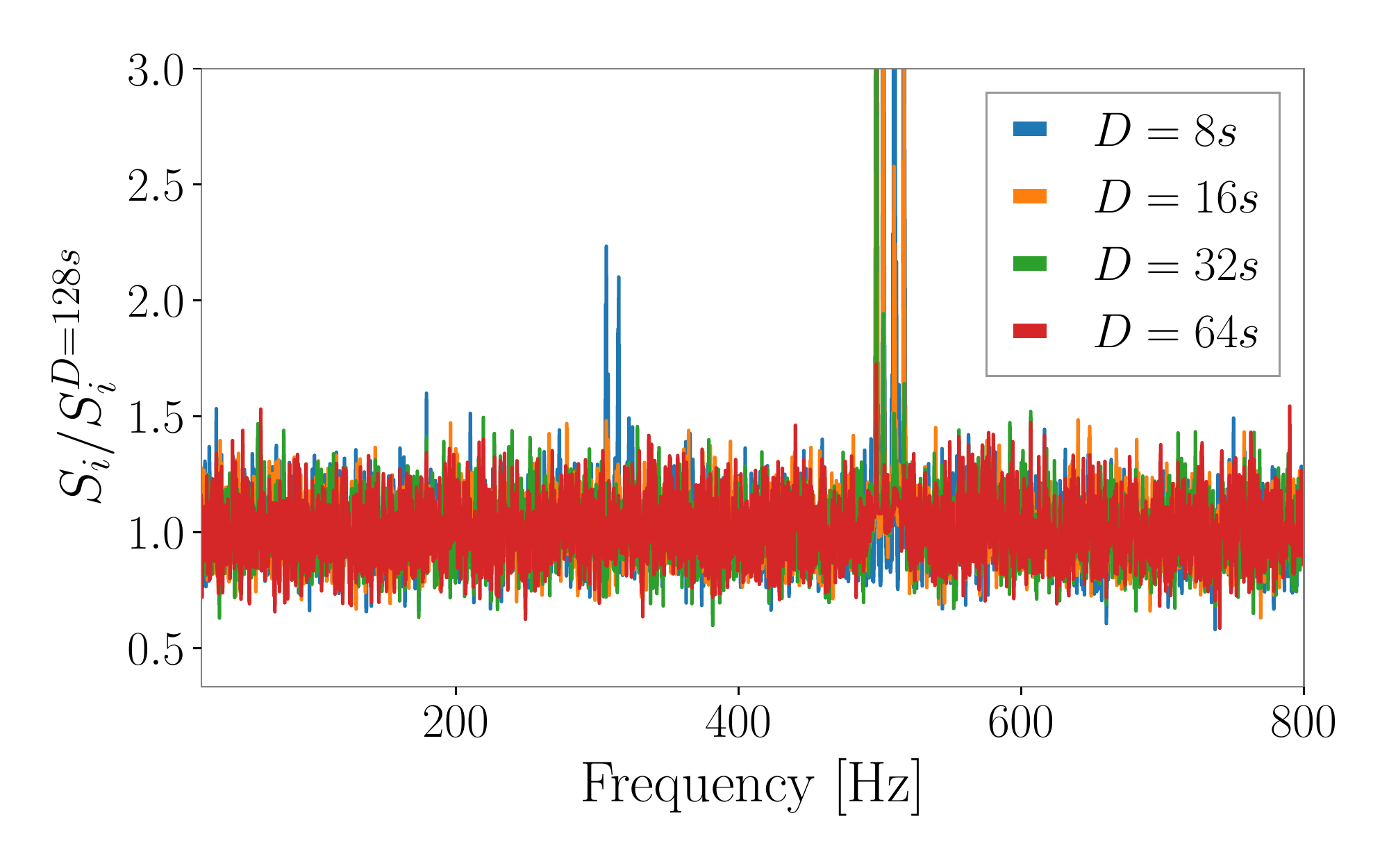}
    \caption{
    The ratio of the inferred finite duration \ac{PSD} ($S_{i}$) with a $\unit[1 / 4]{Hz}$ resolution when using different longer segment durations ($D$) to estimate the infinite-duration \ac{PSD} ($\mathcal{S}_{\mu}$).
    }
    \label{fig:psd-ratio-duration}
\end{figure}

To assess the convergence of this method, we compute $S_{i}$ using a range of values of $D$.
In Fig.~\ref{fig:psd-ratio-duration}, we show the ratio of the finite-duration \ac{PSD}s to the finite-duration \ac{PSD} obtained when using $D=\unit[128]{s}$ (the longest duration we consider).
The estimate quickly converges with increasing $D$ away from the spectral lines.
Close to the forest of large lines around $\unit[500]{Hz}$, the difference between the $D=\unit[64]{s}$ and $D=\unit[128]{s}$ estimates is $\approx 20\%$.
We therefore infer that $D=\unit[64]{s}$ is sufficiently converged.
The key quantity for ensuring adequate convergence is the ratio between $D$ and the original segment duration.
In this case, $\mathcal{S}_{\mu}$ has a resolution $16\times$ as fine as $S_{i}$.
We use this procedure when analyzing real data throughout this paper.

\begin{algorithm*}
\caption{Computing the regularized inverse PSD matrix from real data}
\SetAlgoLined
    \KwResult{regularized inverse PSD matrix (Equation~\ref{eq:regularized-inverse-covariance}) }
    \begin{enumerate}
        \item load $\unit[512]{s}$ of data ending $\unit[2]{s}$ before the analysis segment begins\;
        \item divide into $4 \times \unit[128]{s}$ chunks\;
        \item apply a Hann window (Tukey $\alpha=1$) to each chunk and FFT\;
        \item take a median average of power in each chunk to generate the ``infinite''-duration \ac{PSD} ($\mathcal{S}_{\mu}$)\;
        \item define the infinite-duration window ($w_\psi$) as a $\unit[128]{s}$ time series according to Equation~\ref{eq:inifite-tukey}\;
        \item compute the finite-duration covariance matrix ($C_{\ij}$) using Equation~\ref{eq:exact-covariance}\;
        \item compute the \ac{SVD} (Equation~\ref{eq:svd}) and regularized inverse ($\bar{C}_{ij}$) as outlined in Section~\ref{sec:inversion}\;
    \end{enumerate}
\label{algo:psd-calculation}
\end{algorithm*}

\section{Computing the finite-duration covariance matrix}\label{computing}

For a typical $\unit[4]{s}$ data segment, with sampling frequency $\unit[2048]{Hz}$, analyzed with a $\unit[1 / 128]{Hz}$ noise model, we must perform the matrix operations in Equation~\ref{eq:exact-covariance} with $\mathcal{O}(2^{18}\times 2^{18})$ elements.
A naive implementation at double precision would require a prohibitive amount of computational resources.
Fortunately, the computation can be performed much more computationally efficiently.
The first thing we note is that $\mathcal{C}_{\mu\nu}$ is diagonal and $\tilde{W}_{\mu\nu}$ is a circulant matrix, i.e., it is fully specified by a single row/column ($\tilde{w}_{\mu}$).
Since each matrix can be represented using a single vector, we do not need to form any matrices with the $\unit[1 / 128]{Hz}$ resolution, dramatically reducing memory requirements.
We also identify that $C_{ij}$ is a Hermitian matrix, reducing the computational cost by a factor of two.

We provide functions to compute the coarsened PSD matrix from a frequency-domain window and PSD implemented in \texttt{Cython} and \texttt{cupy} compatible \texttt{CUDA}.
The former runs in $\mathcal{O}(N^3)$ time and the latter in $\mathcal{O}(N)$ wall time.
The latter is used to produce the results in this paper and is less computationally expensive than the \ac{SVD} performed on the coarsened data.

In Algorithm~\ref{algo:psd-calculation}, we describe the process used to compute the regularized inverse covariance matrices used for our binary black hole analyses.
We note that the method presented here requires an estimate for the true underlying power spectral density.
In practice, we estimate this from the data by taking a median average of the PSD in several longer segments.

\section{Additional figures}\label{appendix:event-plots}

\begin{figure*}
    \centering
    \includegraphics[width=\linewidth]{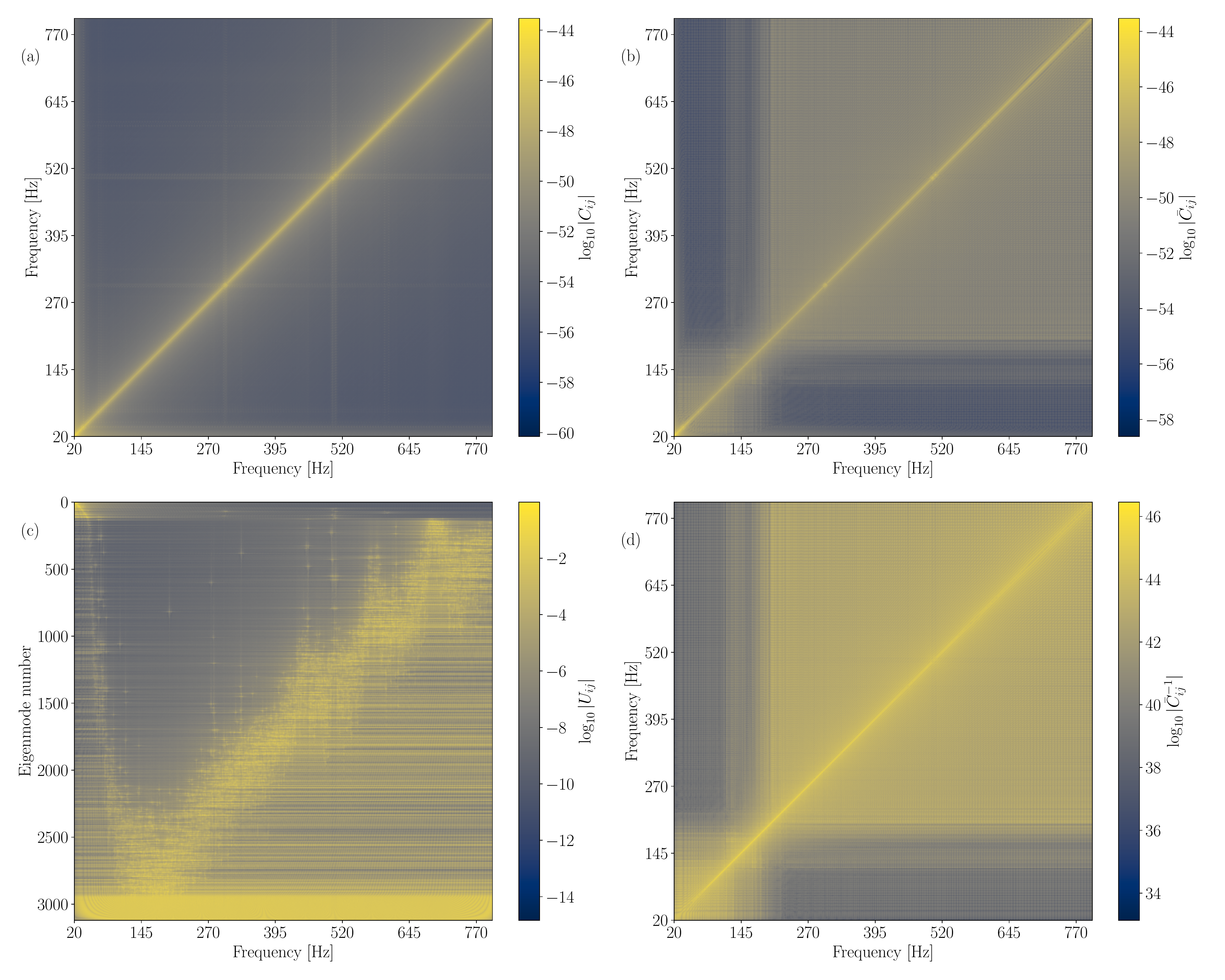}
    \caption{
    \ac{PSD} matrix (top left), regularized \ac{PSD} matrix (top right), \ac{SVD} eigenmatrix (bottom left), and regularized inverse \ac{PSD} matrix (bottom right) for the LIGO Hanford observatory at the time of GW170814.
    }
    \label{fig:GW170814-H1-matrices}
\end{figure*}

\begin{figure*}
    \centering
    \includegraphics[width=\linewidth]{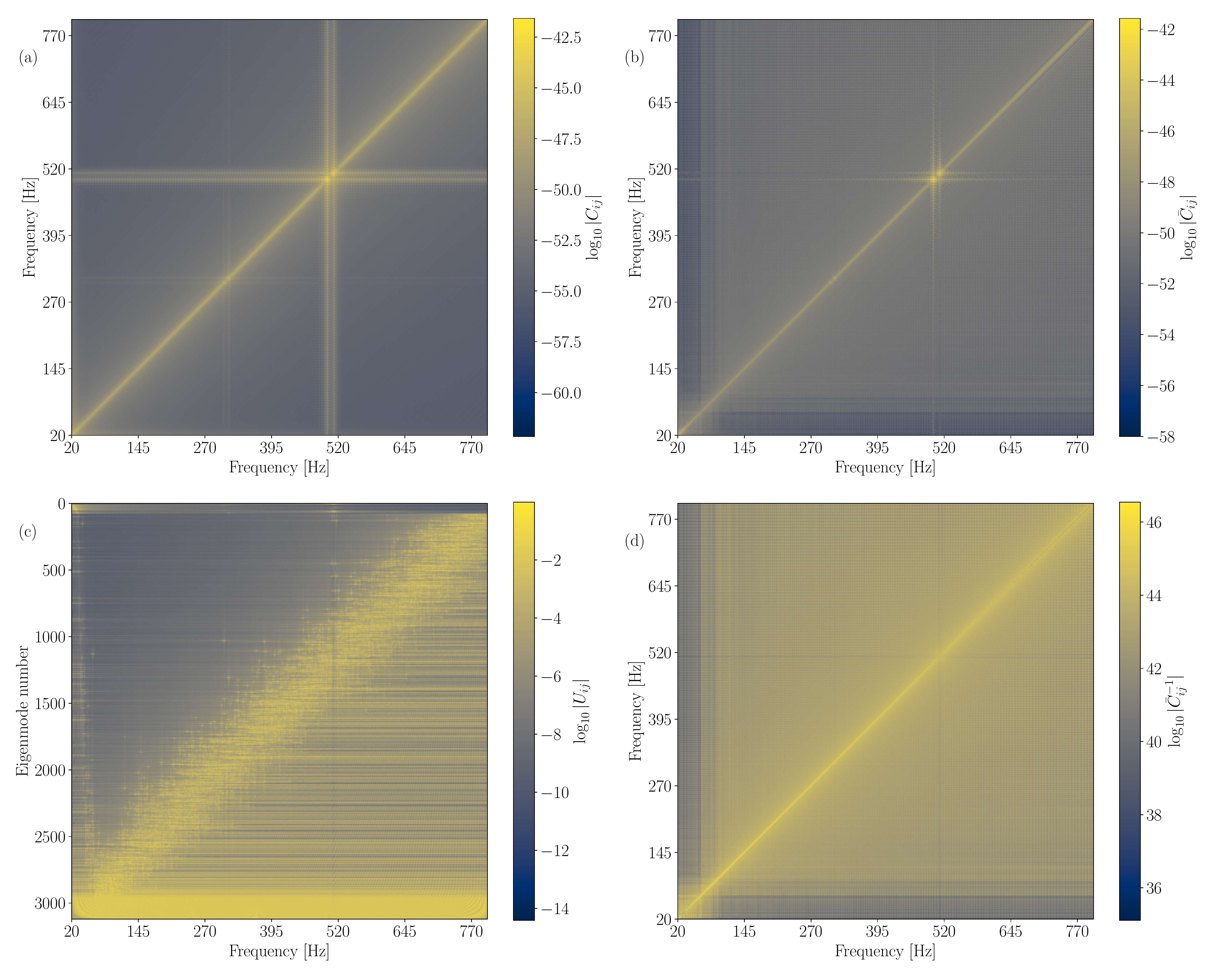}
    \caption{
    \ac{PSD} matrix (top left), regularized \ac{PSD} matrix (top right), \ac{SVD} eigenmatrix (bottom left), and regularized inverse \ac{PSD} matrix (bottom right) for the LIGO Livingston observatory at the time of GW170814.
    }
    \label{fig:GW170814-L1-matrices}
\end{figure*}

\begin{figure*}
    \centering
    \includegraphics[width=\linewidth]{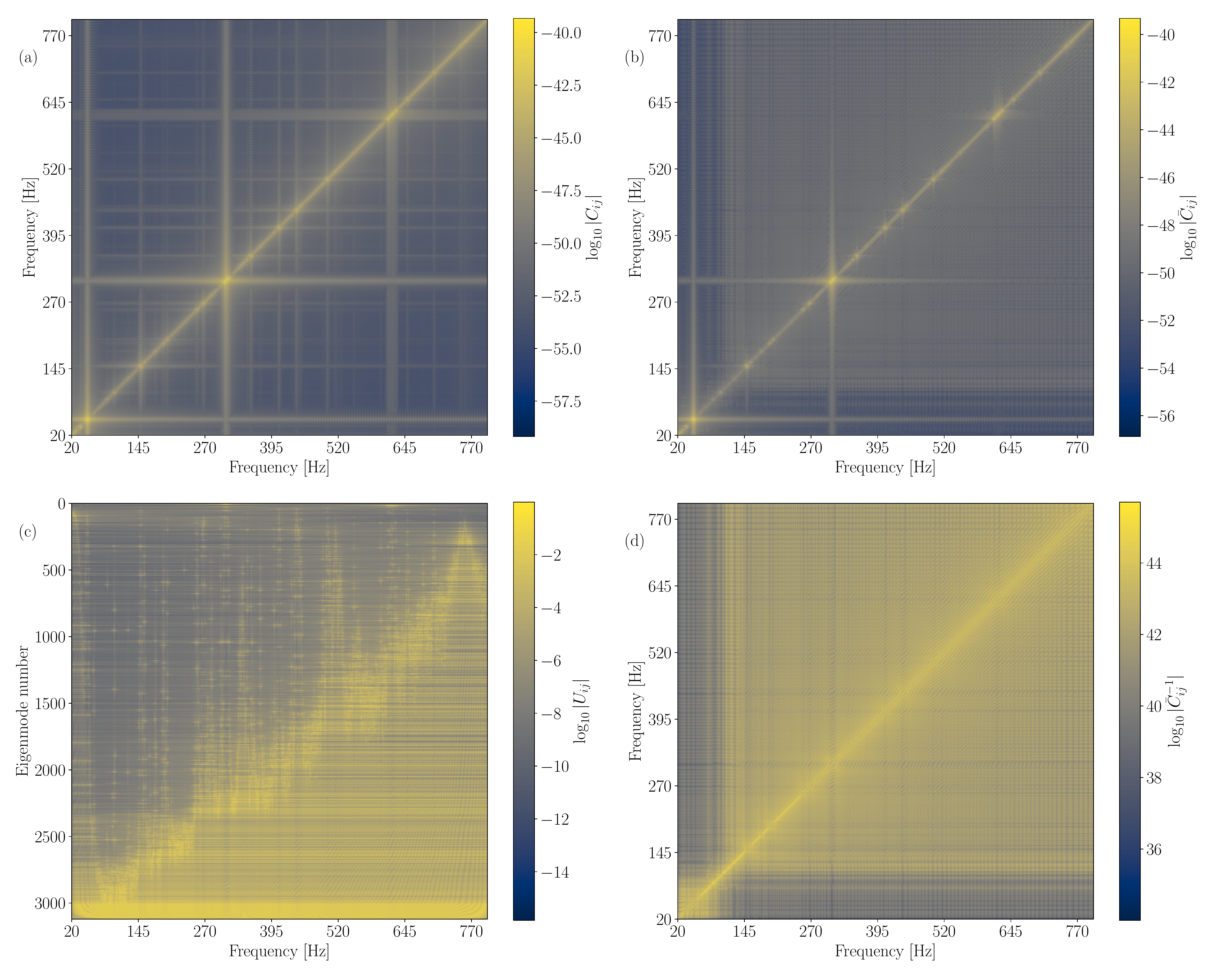}
    \caption{
    \ac{PSD} matrix (top left), regularized \ac{PSD} matrix (top right), \ac{SVD} eigenmatrix (bottom left), and regularized inverse \ac{PSD} matrix (bottom right) for the Virgo observatory at the time of GW170814.
    }
    \label{fig:GW170814-V1-matrices}
\end{figure*}

In this Appendix we show the \ac{PSD} matrix (top left), regularized \ac{PSD} matrix (top right), \ac{SVD} eigenmatrix (bottom left), and regularized inverse \ac{PSD} matrix (bottom right) for our analysis of GW170814 for LIGO Hanford (Fig.~\ref{fig:GW170814-H1-matrices}), LIGO Livingston (Fig.~\ref{fig:GW170814-L1-matrices}), and Virgo (Fig.~\ref{fig:GW170814-V1-matrices}).
We note that the data for all three interferometers show the same qualitative features and quantitative differences determined by the specific sensitivity of each interferometer.

We see that the \ac{PSD} matrix is dominated by the leading diagonal and nearby frequencies and correlations at frequencies corresponding to spectral lines.
The correlations from the spectral lines are less pronounced in the regularized \ac{PSD} matrix, however, there is more broadband correlation between frequencies above/below the most sensitive frequency.
The divide between frequencies above and below the most sensitive frequency can also be seen in the \ac{SVD} and regularized inverse \ac{PSD} matrices.

\section{Window overlaps}\label{app:window-overlaps}

\begin{figure}
    \centering
    \includegraphics[width=\linewidth]{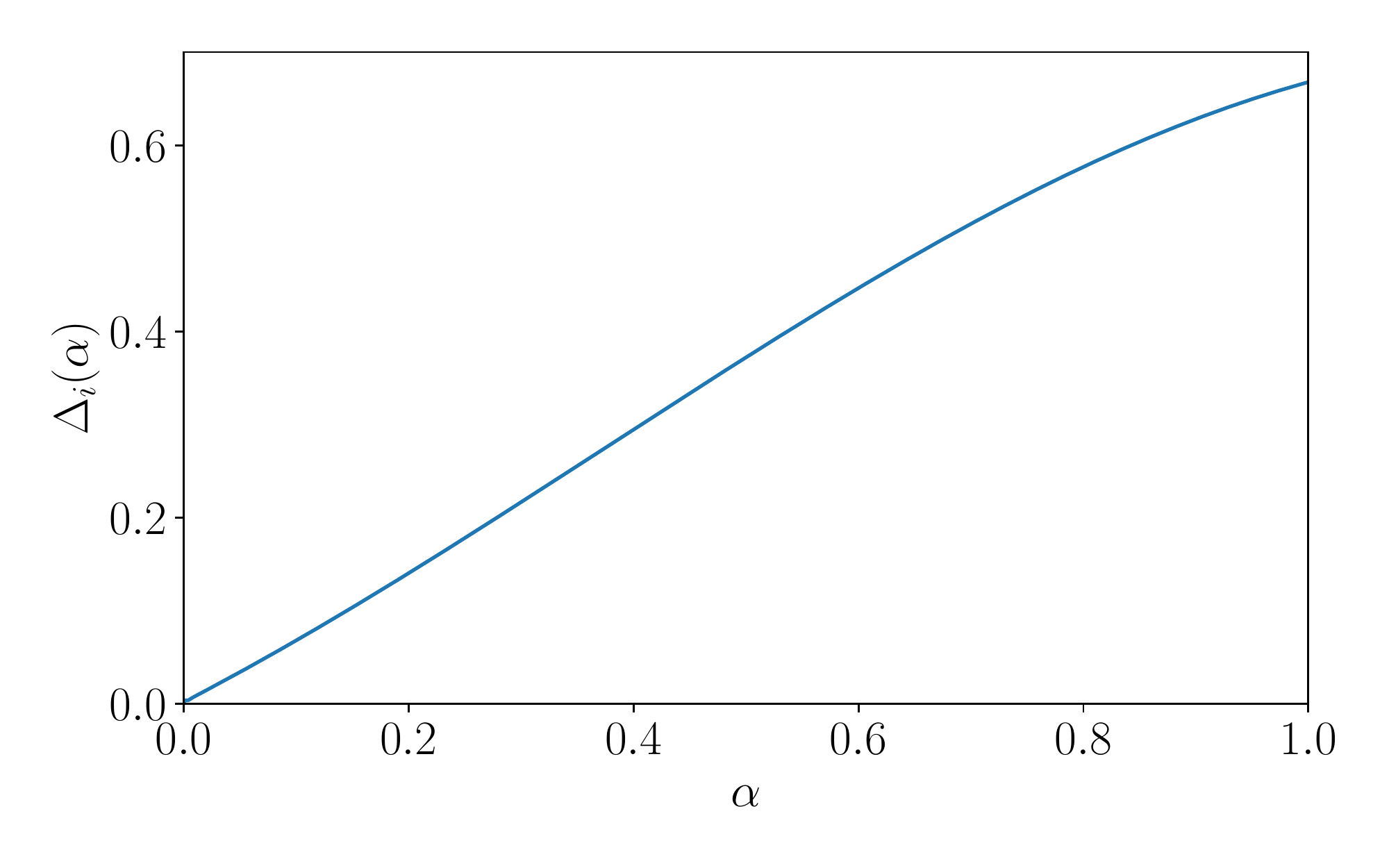}
    \caption{
    Fractional contamination for white noise (Equation~\ref{eq:leakage-white}) as a function of Tukey $\alpha$ parameter.
    We note that the bias is a monotonic function of $\alpha$ ranging from $\Delta_{i} = 0$ for $\alpha = 0$ to $\Delta_{i} \approx 2 / 3$ for $\alpha = 1$.
    }
    \label{fig:delta}
\end{figure}

In order to quantify spectral leakage for white noise we find Equation~\ref{eq:leakage-white}:
\begin{equation}
    \Delta_{i} = \frac{\max_{i \neq j} \left| \sum_{\mu} \tilde{w}_{i - \mu} \tilde{w}^{*}_{j - \mu} \right|}{\left| \sum_{\mu} \tilde{w}_{i - \mu} \tilde{w}^{*}_{i - \mu} \right|}.
\end{equation}
The denominator in this expression is simply the total window power $\overline{w^2}$.
In the continuum limit, the numerator can be written
\begin{equation}
    \Xi(n) = \int^{\infty}_{-\infty} df \tilde{w}(f) \tilde{w}^{*}(f + n / T) \quad (n \in \mathbb{Z}).
\end{equation}
Here $n = i - j$ and the denominator of Equation~\ref{eq:leakage-white} is $\Xi(n=0)$.

For the rectangular and Hann windows, the frequency-domain representations of the windows are
\begin{align}
    \tilde{w}(f; \alpha=0) &= \frac{\sin(\pi f T)}{\pi f} \\
    \tilde{w}(f; \alpha=1) &= \frac{\sin(\pi f T)}{2 \pi f (1 - T^2f^2)},
\end{align}
and we find
\begin{align}
    \Xi(n; \alpha=0) &= \frac{T \sin\left(\pi n \right)}{4 \pi n} \\
    &= \begin{cases} 
        \frac{T}{4} & n = 0 \\
        0 & {\rm else}
    \end{cases}
\end{align}
\begin{align}
    \Xi(n; \alpha=1) &= \frac{3 T \sin\left(\pi n \right)}{2 \pi n(n^2 - 1)(n^2 - 4)} \\
    &= \begin{cases}
        \frac{3 T}{8} & n = 0 \\
        \frac{T}{4} & |n| = 1 \\
        \frac{T}{16} & |n| = 2 \\
        0 & {\rm else}
    \end{cases}.
\end{align}
This trivially shows us that for a rectangular window $\Delta_{i}(\alpha=0) = 0$ and $\Delta_{i}(\alpha=1) = 2 / 3$ in the continuum limit.
For finite duration window functions, the rectangular window still gives $\Delta_{i}(\alpha=0) = 0$ and the Hann window still has a maximum for $|n| = 1$.
We note that an equivalent result is shown in~\cite{Brillinger2001, Lahiri2003} in the discussion of asymptotic independence.

The generic Tukey window does not have an analytic Fourier transform, however, numerical experiments confirm that for all other values of the $\alpha$ parameter, neighboring frequency bins are not independent and $\Delta_{i}$ is a monotonically increasing function of $\alpha$.
In Figure~\ref{fig:delta}, we show $\Delta_{i}$ as a function of $\alpha$.

\bibliography{refs}

\end{document}